# Theory of superlensing with complex frequency illuminations

P. Lalanne[1*], T. Wu

[1] Laboratoire Photonique, Numérique et Nanosciences (LP2N), IOGS- Université de Bordeaux-CNRS, 33400 Talence cedex, France

* philippe.lalanne@institutoptique.fr

**Abstract.** Recent experiments have demonstrated that the resolution of superlensing slabs can be significantly enhanced with complex frequency illuminations. In this study, we introduce a novel theoretical framework for analyzing superlensing. The framework offers new and transparent insights. It helps clarify what resolution can be expected with complex frequency, or more generally pulse illuminations, but it also highlights inherent limitations and tempers high expectations raised by the recent electromagnetic experiments.

Recently, excitations that oscillate at complex-valued frequency have been implemented in electromagnetics or acoustics with tailored waveforms whose amplitudes grow or decay exponentially in time. This advancement has led to demonstrations of new phenomena that change our way of thinking passive linear systems [1].

Notable examples include manipulation of scattering and extinction cross sections beyond passive limits [2-4], enhanced sensing by effectively increasing the Q factors of nanoresonators [5], parity-time symmetry transitions without active elements [3], subwavelength focusing [6]. It has also been demonstrated that the resolution of superlenses [7,8] can be significantly enhanced when illuminated by a monochromatic wave that decays exponentially in time, using acoustic [9] and electromagnetic [10] waves, reviving the long-sought “holy grail” of surpassing the conventional resolution limit.

Since superlenses require large field intensities to compensate the exponential decays of evanescent waves, absorption is a key factor limiting their performance. Thus the experiments in [9-10] have naturally focused on complex frequencies providing "virtual gain" using the transformation $\omega \rightarrow \omega - i\gamma/2$, which enforces the imaginary component of the complex frequency and the superlens material loss to be matched, leading to $\mathrm{Im}(\varepsilon) = 0$ [1,9-11].

The virtual gain approach is primarily based on intuition. A theoretical framework that transparently elucidates the role of complex-frequency illuminations [9-10], or more generally pulse illuminations [12], in superlensing imaging is still absent.

Hereafter, we introduce quasinormal mode (QNM) theory [13] within the framework of superlensing, offering several key insights. Unlike classical models, it highlights the surface modes of superlenses, providing clarity on previously ambiguous discussions [12,14] about the role of backbending in surface-mode dispersion curves on the resolution. Our theory also stresses the fact that optimal performance does not necessarily occur at $\mathrm{Im}(\varepsilon) = 0$ and that both the real and imaginary components of the excitation frequency should be carefully optimized. Additionally, it offers a comprehensive understanding of how to mitigate the contamination of the steady-state response by inevitable transients. Finally, it tempers the high expectations set by recent experiments reporting significant resolution enhancement in the infrared.

## Superlens surface modes.

The inset in Fig. 1 illustrates the superlens geometry. The geometrical and material parameters are directly inspired from the infrared study in [10]. The lens has a thickness $d$ and a frequency-dependent permittivity $\varepsilon_2(\omega)$. We model $\varepsilon_2(\omega)$ using a Drude dispersion relation, $\varepsilon_2(\omega)/\varepsilon_\infty = 1 - \frac{\omega_p^2}{\omega^2 + i\omega\gamma}$, with parameters fitted to match the SiC permittivities in the thermal infrared. The surrounding medium has a frequency-independent background permittivity $\varepsilon_1$. All materials are non-magnetic. We adopt a time dependence of $\exp(-i\omega t)$ to study the steady-state regime.

The superlens is illuminated by a source emitting at frequency $\omega$, which may be complex valued. As in [10], the source may represent, for example, the electromagnetic field scattered by an object with deep subwavelength features. Immediately after the object, the field can be expressed as a plane wave expansion, i.e., a Fourier integral over the in-plane spatial frequencies. For simplicity, and without loss of generality, we assume that the $y$-component $k_y$ of the in-plane wavevector $\mathbf{k}_{||} = [k_x, k_y]$ is null, so the field varies only along the $x$-direction. Importantly, $k_x$ remains real, regardless of whether the excitation frequency $\omega$ is real or complex.

Thus, the superlens surface modes have complex-valued frequencies and real $k_x$. They are found by looking for the solutions of a transcendental equation [7,8,15]

$$\tilde{u} = \exp\left(i\tilde{k}_{2z}d\right) = \pm\frac{\varepsilon_2(\tilde{\omega})\tilde{k}_{1z}+\varepsilon_1\tilde{k}_{2z}}{\varepsilon_2(\tilde{\omega})\tilde{k}_{1z}-\varepsilon_1\tilde{k}_{2z}}. \quad (1)$$

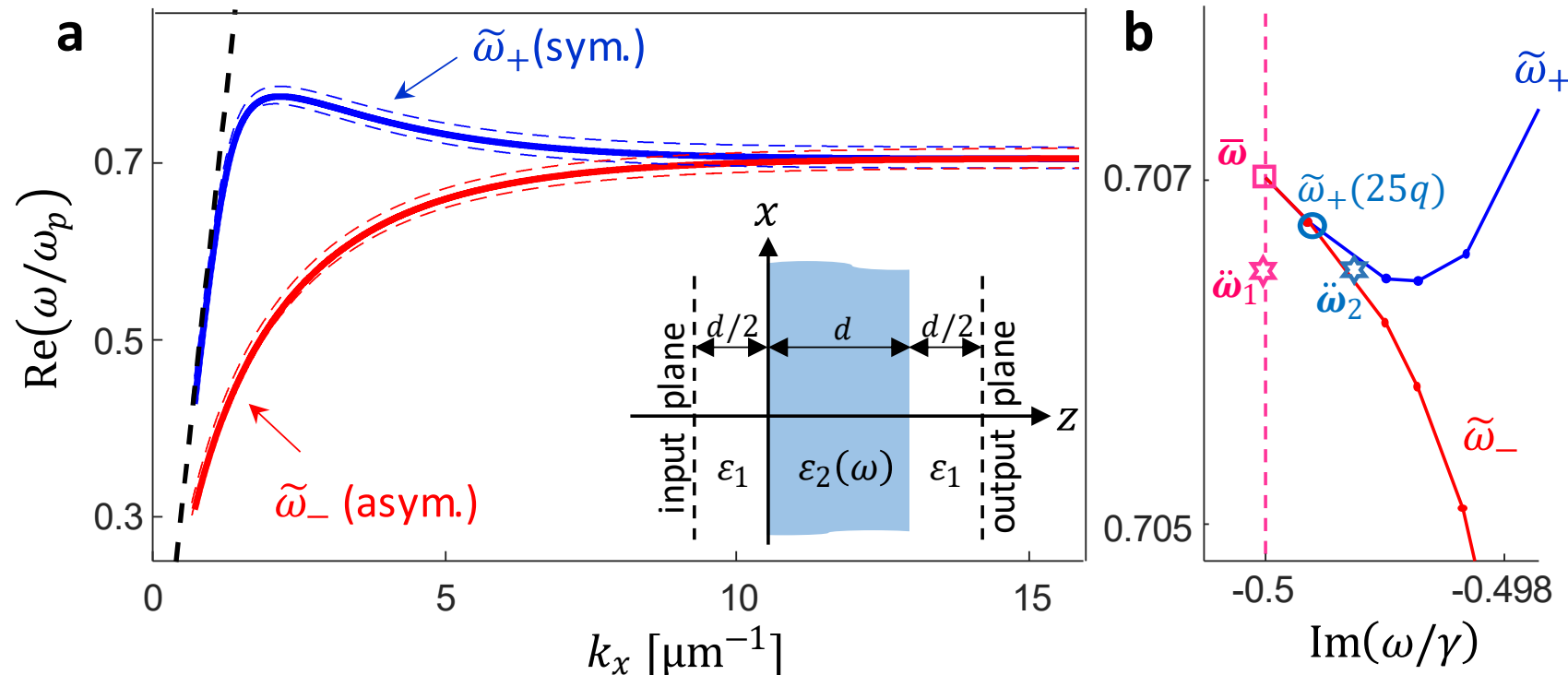

FIG. 1. (a) Dispersion relation of the two polaritonic QNM frequencies, $\tilde{\omega}_+(k_x)$ and $\tilde{\omega}_-(k_x)$, which are dominant below the plasma frequency. Dashed curves correspond to $\mathrm{Re}(\tilde{\omega}) \pm \mathrm{Im}(\tilde{\omega})$. The dashed line is the silica light line. The upper branch frequency $\tilde{\omega}_+(k_x)$ corresponds to a symmetric field: $\tilde{H}_y(x, z - d/2) = \tilde{H}_y(x, -z + d/2)$. The lower branch mode is antisymmetric. Inset shows the superlens geometry. (b) Loci in the complex-frequency plane $\mathrm{Re}(\omega/\omega_p)$ vs $\mathrm{Im}(\omega/\gamma)$ of four key frequencies used in this work to optimize the lens performance. The two fuchsia marks and the dashed fuchsia line correspond to $\mathrm{Im}(\omega) = -i\gamma/2$. $\bar{\omega}$ is the resonant frequency determined by the condition $\varepsilon_2(\bar{\omega}) = -\varepsilon_1$. $\ddot{\omega}_1$ and $\ddot{\omega}_2$ are optimized frequencies. The Drude parameters are $\varepsilon_\infty = \varepsilon_1 = 4$, $\omega_p = 2.42 \times 10^{14}$ rad/s, and $\gamma = 0.056 \times 10^{14}$ rad/s.

Hereafter, tildes are used to denote quantities related to the resonance modes, which will be referred to as QNMs hereafter. For instance, $\tilde{\omega}$ denotes the QNM eigenfrequency, whereas $\omega$ denotes the frequency of the incident plane wave, even if this frequency is complex valued. Consistently, in Eq. (1), we have $\tilde{k}_{1z} = \left(\varepsilon_1\tilde{k}^2 - k_x^2\right)^{1/2}$ and $\tilde{k}_{2z} = \left(\varepsilon_2(\tilde{\omega})\tilde{k}^2 - k_x^2\right)^{1/2}$, with $\tilde{k} = \tilde{\omega}/c$. These expressions require a definition of a branch cut for the square root. We choose the sign of $\sqrt{x}$ such that $Re\left(\sqrt{x}\right) + Im\left(\sqrt{x}\right) > 0$ [16].

The transcendental Eq. (1) admits an *infinity* of solutions with complex-valued frequencies for every $k_x$ [12,17]. Most of these solutions have resonant frequencies larger than the plasma frequency $\omega_p$. Below $\omega_p$, there are only two dominant QNMs with frequencies $\tilde{\omega}_+(k_x)$ and $\tilde{\omega}_-(k_x)$ [18], where the sign '+' ('−') holds for the (anti-) symmetric mode. Figure 1 shows their dispersion relations, computed numerically by solving the transcendental Eq. (1), see Suppl. Note 2.2. For large $k_x$, the surface modes of two interfaces do not interact, and the two polaritonic branches become nearly degenerate at frequency $\bar{\omega}$. These are well known results [18,15].

## Steady-state QNM transfer function.

To derive the new transmission formula, we follow the QNM formalism developed in [19]. For a mode with in-plane wavevector $k_x$, the QNM normalization inherently involves the mode with opposite wavevector $-k_x$. Introducing the convenient notations $\tilde{\mathbf{H}}(x, z, k_x) = \tilde{\mathbf{h}}(z, k_x)\exp(ik_x x)$ and $\tilde{\mathbf{E}}(x, z, k_x) = \tilde{\mathbf{e}}(z, k_x)\exp(ik_x x)$, the normalized fields satisfy the condition

$$1 = \int_{-\infty}^{\infty}\left(\frac{\partial\omega\varepsilon}{\partial\omega}\tilde{\mathbf{e}}(z, k_x)\cdot\tilde{\mathbf{e}}(z, -k_x) - \mu_0\tilde{\mathbf{h}}(z, k_x)\cdot\tilde{\mathbf{h}}(z, -k_x)\right)dz, \quad (2)$$

with $\tilde{\mathbf{h}}(z,k_x)=\tilde{\mathbf{h}}(z,-k_x)$, $\tilde{\mathbf{e}}(z,k_x)\cdot\hat{\mathbf{x}}=\tilde{\mathbf{e}}(z,-k_x)\cdot\hat{\mathbf{x}}$ and $\tilde{\mathbf{e}}(z,k_x)\cdot\hat{\mathbf{z}}=-\tilde{\mathbf{e}}(z,-k_x)\cdot\hat{\mathbf{z}}$. Throughout this work, all QNMs are assumed to be normalized, and the notations $\tilde{\mathbf{E}},\tilde{\mathbf{e}},\tilde{\mathbf{H}}$ or $\tilde{\mathbf{h}}$ refer to these normalized fields.

The superlens is illuminated by a p-polarized plane wave at frequency $\omega$ and in-plane wavevector $k_x$, with magnetic field $\mathbf{H}_{inc}(x,z,k_x)=H_{inc}\exp[i(k_{1z}z+k_x x)]\,\hat{\mathbf{y}}$. To express the fields scattered by the superlens, $\mathbf{E}_S(\mathbf{r},\omega,k_x)$ and $\mathbf{H}_S(\mathbf{r},\omega,k_x)$, in the QNM basis, we use the orthogonality-decomposition method and introduce the auxiliary polarizabilities and currents: $\mathbf{P}_t(x,z,k_x)=[\varepsilon_2(\omega)-\varepsilon_\infty][\mathbf{E}_S(x,z,k_x)+\mathbf{E}_{inc}(x,z,k_x)]$ and $\mathbf{J}_t(x,z,k_x)=-i\omega\mathbf{P}_t(x,z,k_x)$ [19].

The scattered fields, augmented by the auxiliary fields, admit a unique QNM expansion that is complete inside the superlens layer [19]

$$\begin{bmatrix}\mathbf{E}_S(\mathbf{r},\omega,k_x)\\ \mathbf{H}_S(\mathbf{r},\omega,k_x)\\ \mathbf{P}_t(\mathbf{r},\omega,k_x)\\ \mathbf{J}_t(\mathbf{r},\omega,k_x)\end{bmatrix}=\sum_{m=1,2\ldots}\alpha_m(\omega,k_x)\begin{bmatrix}\tilde{\mathbf{E}}_m(\mathbf{r},k_x)\\ \tilde{\mathbf{H}}_m(\mathbf{r},k_x)\\ \tilde{\mathbf{P}}_m(\mathbf{r},k_x)\\ \tilde{\mathbf{J}}_m(\mathbf{r},k_x)\end{bmatrix}. \tag{3}$$

Each mode $m$ combines electromagnetic and material (polarization) contributions into a quadrivector representation. The auxiliary fields are defined as $\tilde{\mathbf{P}}_m=[\varepsilon_2(\tilde{\omega}_m)-\varepsilon_\infty]\tilde{\mathbf{E}}_m$ and $\tilde{\mathbf{J}}_m=-i\tilde{\omega}_m\tilde{\mathbf{P}}_m$.

The modal excitation coefficients in Eq. (3) are given by the overlap between the QNM electric field and the electric field $\mathbf{E}_{inc}$ of the incident plane wave

$$\alpha_m(\omega,k_x)=\varepsilon_0\left[\frac{\tilde{\omega}_m}{\tilde{\omega}_m-\omega}(\varepsilon_2(\tilde{\omega}_m)-\varepsilon_1)+(\varepsilon_1-\varepsilon_\infty)\right]\int_0^d\tilde{\mathbf{E}}_m(x,z,-k_x)\cdot\mathbf{E}_{inc}\,dz. \tag{4}$$

We define the transfer function $t(k_x,\omega)$ between the planes $z=-d/2$ and $z=3d/2$ as the ratio of the total magnetic field $H_t(x,z,\omega)$ evaluated at these planes, $t=H_t(x,3d/2,\omega)/H_t(x,-d/2,\omega)$, where the total magnetic field is the sum of the incident field and the scattered field: $H_t(\mathbf{r},\omega,k_x)=H_S(\mathbf{r},\omega,k_x)+H_{inc}(\mathbf{r},\omega,k_x)$. Since the QNM electromagnetic fields are known analytically (Suppl. Note 2) for all QNMs, an exact analytical expression for $t(k_x,\omega)$ can be derived using Eqs. (2-4).

We now introduce the *only approximation* employed in this work. We assume that the scattered field can be accurately reconstructed by considering only the two polaritonic QNMs shown in Fig. 1: $\mathbf{H}_S(\mathbf{r},\omega,k_x)=\alpha_+(\omega,k_x)\tilde{\mathbf{H}}_+(\mathbf{r},k_x)+\alpha_-(\omega,k_x)\tilde{\mathbf{H}}_-(\mathbf{r},k_x)$. As we will show, this approximation is well justified. The numerical modes contribute negligibly to the near field of the lens, and the Fabry–Perot QNMs have much higher frequencies (i.e., $|\tilde{\omega}_m|\gg\omega_p$), lying well outside the spectral range of interest.

The two-QNM assumption delivers a very compact formula for the transfer function for $\varepsilon_1=\varepsilon_\infty$

$$t(k_x,\omega)\approx\frac{1}{8}\frac{\omega_p}{\omega}\left[\frac{\omega_p}{\tilde{\omega}_--\omega}-\frac{\omega_p}{\tilde{\omega}_+-\omega}\right]\exp\left[-\left(k_x^2-\varepsilon_1\frac{\omega^2}{c^2}\right)^{1/2}d\right], \tag{5}$$

where the important minus sign in the bracket arises from the asymmetry of the two QNMs. Full derivation details are given in Supplementary Section 3. In contrast to the conventional Airy (Fabry–Perot) formula [7-9], Eq. (5) notably highlights the importance of the lens resonances.

To derive the equation, we made, in addition to the main two-QNM approximation, a few minor classical assumptions, specifically that $|k_x|\gg\omega/c$, $|\tilde{\omega}_+|/c$ and $|\tilde{\omega}_-|/c$, letting us expect that Eq. (5) is accurate for large $k_x$.

We have tested the accuracy of Eq. (5) by comparing its predictions with reference data obtained with the 2×2 matrix-transfer formalism [23] with the program given in Supp. Section 3 implemented with the freeware RETICOLOfilm-stack [24]. Many cases have been considered, and systematically we obtained excellent agreement, see a few examples in Figs. S2 and S4. We can thus heavily rely on this equation to analyse the properties of superlenses under complex frequency illuminations.

Equation (5) provides an amazingly simple yet valuable insight into selecting the optimal complex frequency for best performance: specifically, it suggests placing the illumination frequency in the complex plane in a way that optimally excites the two polaritonic dispersion curves. For more complex structures, e.g. multilayered slabs and hyperbolic metamaterials [10,25,26], an expression similar to that of Eq. (5) is anticipated, possibly involving a few additional poles. More broadly, the approach is applicable to arbitrary 3D resonant systems [19].

In an ideal scenario involving lossless materials, the polaritonic dispersion curves closely resemble those in Fig. 1, but in this case, $\tilde{\omega}_+(k_x)$ and $\tilde{\omega}_-(k_x)$ are real valued. Complex frequency illuminations do not implement this "ideal" scenario; they can only approach it. This can be realized by rewriting the denominators in Eq. (5) as $\omega_p/\left(\mathrm{Re}(\tilde{\omega}_\mp - \omega) + i\mathrm{Im}(\tilde{\omega}_\mp - \omega)\right)$. $\mathrm{Im}(\tilde{\omega}_\mp - \omega)$ can be nullified only for a single pole and a single prescribed $k_x$. Therefore, no matter how we chose $\mathrm{Im}(\omega)$, the system will never perfectly mimic an ideal system without loss. In fact, the transformation $\omega \to \omega - i\gamma/2$ implements an artificial medium that amplifies waves via both its effective permittivity and permeability (Suppl. Section 4). As a result, significant discrepancies between the transfer functions computed using $\omega \to \omega - i\gamma/2$ or $\gamma = 0$ generally emerge, see Fig. S4.

The literature on superlensing also introduces some confusion regarding different surface-polariton dispersion relations, $\mathrm{Re}(\tilde{k}_x)$ versus $\omega$ or $\mathrm{Re}(\tilde{\omega})$ versus $k_x$ [27], which provide restricted or unrestricted access to large $k_x$'s, depending on whether backbending is present. For example, it has been argued—based on the incorrect intuition that monochromatic illumination only accesses guided-mode dispersion curves, $\mathrm{Re}(\tilde{k}_x)$ versus $\omega$, which exhibit backbending—that the resolution of superlenses could be improved by using time-dependent incident pulses [12]. These pulses, it was argued, would provide access to the QNM dispersion curves shown in Fig. 1a, i.e., $\mathrm{Re}(\tilde{\omega})$ versus $k_x$, without backbending. More recently, it was similarly suggested that removing material losses—by introducing virtual gain—would enable a true paradigm shift, eliminating backbending and providing access to large $k_x$'s [14].

The present approach clarifies the confusion. Regardless of the type of illumination, monochromatic, complex-frequency, or time-dependent pulse, it is always the QNMs that are excited, and access to large $k_x$'s is possible even with significant material loss. The key difference lies in the excitation rate of the QNMs, not in the availability of large $k_x$'s. Therefore, backbending of guided-mode dispersion curves has no relevance to superlensing.

## Resolution improvement with complex frequencies.

Equation (5) also provides new insights into strategies for enhancing superlens resolution by optimally exciting the resonances of the superlens. Figure 2 summarizes the results of our investigations for the superlens geometry examined in [10] at infrared frequencies.

To better visualize the results, we introduce the parameter $q = 2\pi\sqrt{\varepsilon_1}/\lambda_0$ to serve as a normalization factor equal to the wavector modulus in silica at wavelength $\lambda_0 = 11$ µm. Accordingly, $k_x/q$ can be interpreted as the number of parallel wavevector units in the incident medium. Note the dashed green curve that serves as a reference performance expected at real frequency.

Four strategic $\omega$-values, $\bar{\omega}$, $\tilde{\omega}_+(k_x = 25q)$, $\ddot{\omega}_1$ and $\ddot{\omega}_2$, are considered. They are positioned in the complex frequency plane with respect to the dispersion curves $\tilde{\omega}_+(k_x)$ and $\tilde{\omega}_-(k_x)$ in Fig. 1a with circle, square and star marks. Same marks are consistently used in Fig. 2. To compare their performance, we define the cutoff $K_x$ as the maximum spatial frequency such that transmission exceeds unity for all $q < k_x < K_x$.

The first frequency, $\bar{\omega}$, is the resonant frequency of the surface polariton mode supported by each interface of the superlens. This frequency corresponds to the asymptotic degenerate limits of $\tilde{\omega}_+$ and $\tilde{\omega}_-$ at large $k_x$. It is very natural to consider this frequency as it corresponds to the impedance (or perfect lens) condition, $\varepsilon_2(\bar{\omega}) = -\varepsilon_1$ [7,9]. The corresponding transmission, computed with the 2×2 matrix-transfer formalism, is shown as the fuchsia solid curve in Fig. 2. It remains above unity up to a cutoff of $K_x = 12q$.

The second frequency is the resonant frequency $\omega = \tilde{\omega}_+(25q)$ of the upper branch polariton for a large spatial frequency $k_x = 25q$. The corresponding transmission, also computed with the 2×2 matrix-transfer formalism, is shown as the blue dashed curve. However, the overall performance shows only a marginal improvement, $K_x = 13q$, due to the dominance of the exponential damping term in Eq. (5) for large $k_x$'s. As predicted by Eq. (5), a divergence is observed at $k_x = 25q$. As anticipated from Eq. (5), similar transmission behavior is observed for $\omega = \tilde{\omega}_-(25q)$.

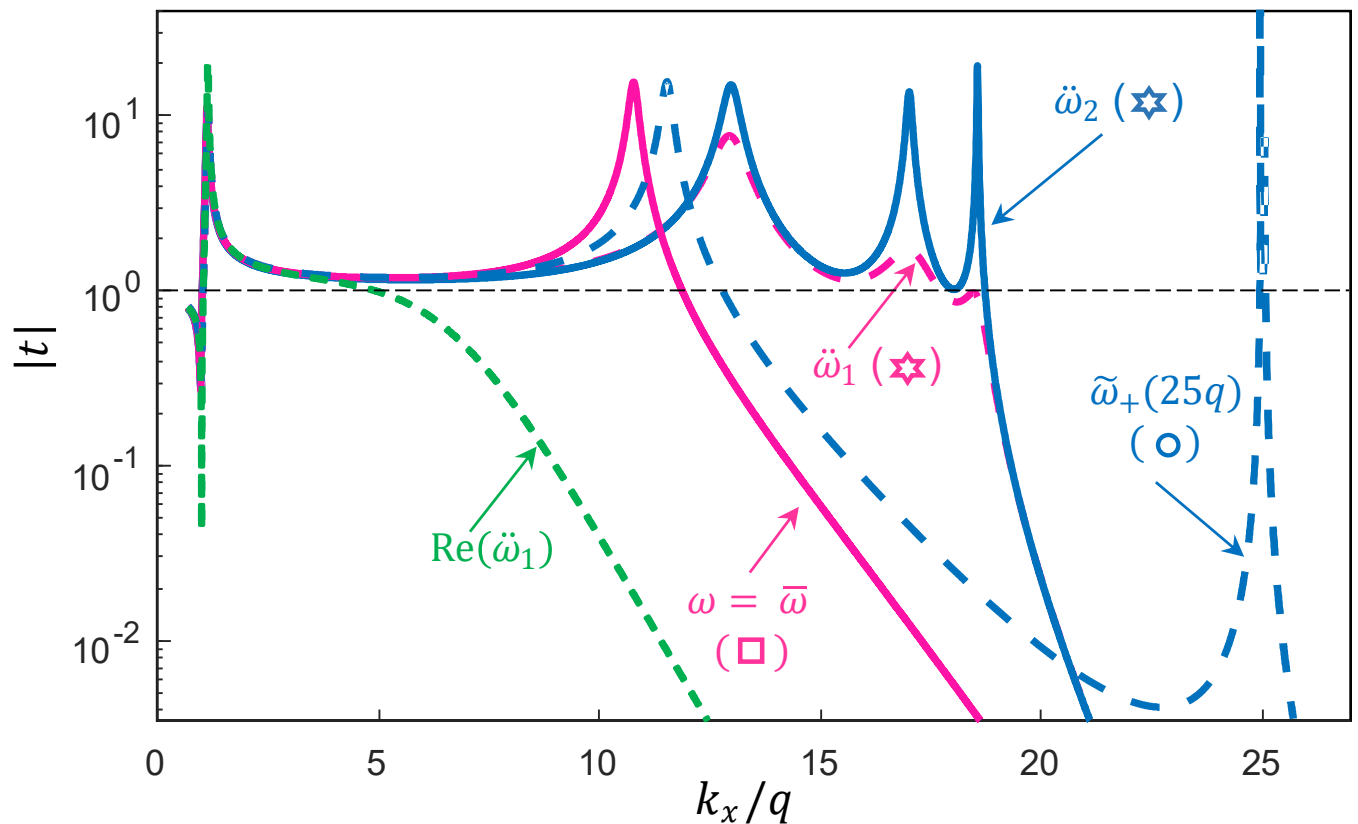

FIG. 2. $|t|$ for various complex-frequency illuminations of interest. Fuchsia solid curve: $\omega = \varpi$ ($\varepsilon_2(\varpi) = -\varepsilon_1$). Blue dashed curve: $\omega = \tilde{\omega}_+(25q)$, a pole yielding divergent transmission at $k_x = 25q$ . Fuchsia dashed curve: optimized transmission obtained for the illumination frequency $\ddot{\omega}_1$ (fuchsia star in Fig. 1b). The optimization is restricted to frequencies $\omega$ such that $\mathrm{Im}(\omega) = -\gamma/2$. Blue solid curve: best performance obtained for the illumination frequency $\ddot{\omega}_2$ (blue star in Fig. 1b). All curves are calculated with the $2 \times 2$ matrix-transfer formalism. The dashed green curved, calculated for $\omega = \mathrm{Re}(\ddot{\omega}_1)$, is a typical reference performance at real frequency.

To compute the other two frequencies, we resort to optimization, looking for complex frequencies that ensure the largest cutoff. We first assume $\mathrm{Im}(\omega) = -\gamma/2$, implying that we scan the complex $\omega$-plane along the vertical dashed line in Fig. 1b. An enlarged cutoff, $K_x = 18q$ (fuchsia dashed curve), is achieved for the frequency $\ddot{\omega}_1$ labelled with the fuchsia star.

We then do not make any restriction on the imaginary component of the illumination frequency, and scan exhaustively the complex plane to identify the frequency that yields the maximum $K_x$. The computation is very fast with Eq. (5). The optimal complex frequency is denoted $\ddot{\omega}_2$ in Fig. 1b. $K_x$ is slightly increased, from $K_x = 18q$ to $19q$.

Compared to the dashed green curve, the cutoff improves twofold when using the plasmon frequency, $\varpi$, and threefold when using the optimized frequencies, $\ddot{\omega}_1$ or $\ddot{\omega}_2$.

## Observability of steady-state responses.

Exponentially damped illuminations always begin at a specific initial timestep and a transient response must also be considered in addition to the steady-state response described by Eq. (5). The important question of a possible contamination of the steady-state response arises [22,28].

The QNM approach is particularly valuable to answer the question. In Suppl. Section 5.1, we apply recent rigorous formulations of resonator dynamics [29] to investigate the observability of the steady-state response in the general case of a resonator with an arbitrary shape and with a permittivity $\boldsymbol{\varepsilon}_2$ in a background permittivity $\boldsymbol{\varepsilon}_1$. We consider a typical incident wave packet of the form $\mathbf{E}_{inc}(\boldsymbol{r}, t) = \mathbf{E}_0\, S(t - r/c) \exp[-i\omega(t - r/c)]$, where $S(t)$ is a slowly varying sigmoid function. Under minimal assumptions, we show that the modal excitation coefficient $\beta_m(t)$ of a resonator placed at the origin is a combination of a steady-state contribution, $\propto \exp(-i\omega t)$ , and a transient contribution, $\propto \exp(-i\tilde{\omega}_m t)$,

$$\beta_m(t) = S(t) \frac{\langle \mathbf{E}_0 | \varepsilon_0 \Delta\boldsymbol{\varepsilon} | \tilde{\mathbf{E}}_m \rangle}{\tilde{\omega}_m - \omega} [\omega \exp(-i\omega t) - \tilde{\omega}_m \exp(-i\tilde{\omega}_m t)], \tag{6}$$

with $\langle \mathbf{E}_0 | \varepsilon_0 \Delta\boldsymbol{\varepsilon} | \tilde{\mathbf{E}}_m \rangle$ the overlap integral between the electric fields of the incident wave packet and the $m^{\mathrm{th}}$ QNM inside the resonator ($\Delta\boldsymbol{\varepsilon} = \boldsymbol{\varepsilon}_2 - \boldsymbol{\varepsilon}_1 \neq \mathbf{0}$) [19].

Equation (6) provides a general expression applicable to any resonant system. It provides key insights on the observability of the steady-state regime, which requires that the temporal damping of the illumination be smaller than the damping of all dominant modes: $|\mathrm{Im}(\omega)| < |\mathrm{Im}(\tilde{\omega}_m)|$, for all $m$.

Another crucial point arises when the illumination frequency is close to a resonance frequency, i.e., $\omega \approx \tilde{\omega}_m$. The minus sign in the bracket of Eq. (6) indicates that the steady-state and transient responses are of equal magnitude but unfortunately out of phase, implying that the steady state can only be observed after very long times (Fig. S5b). For $\omega = \tilde{\omega}_m$ , Eq. (6) simplifies to $\beta_m(t) = i\omega t\, S(t)\, \langle \mathbf{E}_0 | \varepsilon_0 \Delta\boldsymbol{\varepsilon} | \tilde{\mathbf{E}}_m \rangle \exp(-i\omega t)$. The steady-state regime of the dominant QNMs then loses its typical $(\tilde{\omega}_m - \omega)^{-1}$ divergence which makes it predominant for all $t$. Instead, we have a prefactor that scales as $\omega t$ which is dominant at large $t$ only.

In Fig. 1a, the resonances with extended lifetimes have small $k_x$ values, due to reduced confinement. Thus, the steady-state responses of Fig. 2 are only discernible at high spatial frequencies, where $|\mathrm{Im}(\tilde{\omega}_+)|$ and $|\mathrm{Im}(\tilde{\omega}_-)|$ exceed $|\mathrm{Im}(\omega)|$. Figure S5, in Suppl. Section 5.2, investigates the possibility of expanding the discernability towards lower spatial frequencies by reducing the damping of the excitation illumination. We find that this reduction inevitably compromises imaging performance at high spatial frequencies. Attaining superlens imaging with a purely steady-state response using direct illumination thus appears challenging.

## Discussion.

The cutoff increase using complex frequencies is modest: only a two- to threefold improvement is achieved even with carefully optimized frequencies (Fig. 2). This limited gain comes at the cost of pronounced peaks in the transfer functions, which are well known to degrade imaging performance [8]. To access their impact, following [10], we consider a 60-nm-thick gold grating with three slits of varying widths in the unit cell. As shown in Fig. 3a, the near-field distribution 20 nm above the grating surface displays weak fields near the metal and stronger fields at the slit openings. For the computation, we used the Rigorous Coupled Wave Analysis (RCWA) [30] and a normally-incident plane wave with a magnetic field parallel to the slits.

Since RCWA naturally yields a $k_x$-space expansion of the fields, computing $|tE_x|$ is straightforward. The results (Fig. 3b) reveal some degree of resolution enhancement using complex frequencies. However, the improvement is modest: for the widest slit, the patterns computed with $\varpi$ and $\ddot{\omega}_1$ are only slightly sharper than those obtained at real frequency. The contrast is also much weaker than in the experimental data reported in Fig. 4c of [10]. In particular, the numerical profiles exhibit smoother variation at slit edges, lacking the steep features observed experimentally, despite the use of nearly identical parameters—materials, lens thickness, and slit geometries.

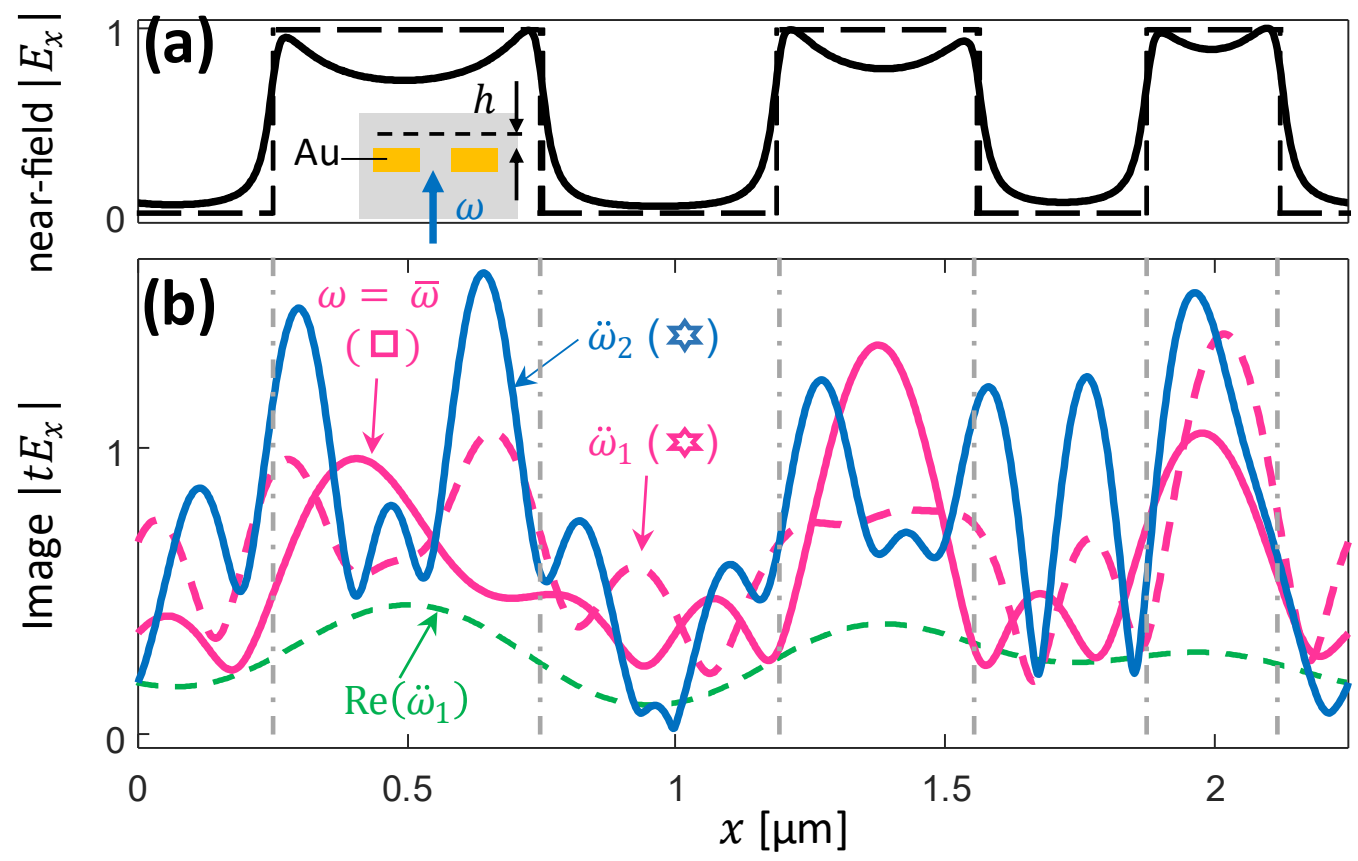

FIG. 3. Imaging performance. (a) Near-field of a 60-nm-thick Au super cell composed of 3 slits with varying widths, 500, 375 and 250 nm. $|E_x(x)|$ is computed at a distance $h = 20$ nm from the grating surface. (b) $|tE_x|$ computed for $\mathrm{Re}(\ddot{\omega}_1)$, $\varpi$, $\ddot{\omega}_1$ and $\ddot{\omega}_2$. These image patterns at both real and complex frequencies resonates with corresponding experimental data in Fig. 4c in [10]. All curves are normalized by the maximum of $|E_x|$. The coordinate $x$ (resp. $z$) is parallel (resp. perpendicular) to the grating surface. Notably, the image patterns $|tE_z|$ for the $z$-component of the electric field (Fig. S8) do not display improved resolution.

Several factors may contribute to this discrepancy. On the theoretical side, the simulation cannot fully capture the tip–near-field interaction, and some uncertainty in the measured quantity is unavoidable [27]. Moreover, the image patterns being extremely sensitive to the illumination frequency—the relative differences between the imaginary and real parts of $\varpi$ and $\ddot{\omega}_2$ are only 0.15% and 0.08%—, even a small uncertainty in identifying the exact complex frequency used in the synthetic reconstruction of [10] can lead to markedly different results. On the experimental side, interferometric near-field reconstructions are highly sensitive to detection noise [27], so potential artifacts cannot be entirely ruled out, even when the experiments are performed with great care.

These findings temper the high expectations raised by recent electromagnetic experiments. The strength of superlenses—their resonance character—is also their limitation, since it prevents flat transfer functions with high cutoffs. For practical applications, e.g. lithography, synthetic reconstructions based on superpositions of real-frequency components should be reconsidered in favor of direct illumination approaches. Moving forward, new strategies—either by harnessing persistent transients or by controlling them with optimized pulse shapes, as suggested in [12]—will be essential for making superlenses a truly practical tool. The present theory is expected to serve as a key step toward these innovations.

*Acknowledgements.* PL thanks Andrea Alù, Seunghwi Kim, Jean-Jacques Greffet, Roman Calpe and Thomas Christopoulos are acknowledged for fruitful discussions.

# Supplementary information: Theory of superlensing with complex frequency illuminations

P. Lalanne[1*], T. Wu

[1] Laboratoire Photonique, Numérique et Nanosciences (LP2N), IOGS- Université de Bordeaux-CNRS, 33400 Talence cedex, France

* philippe.lalanne@institutoptique.fr

## 1. Notations

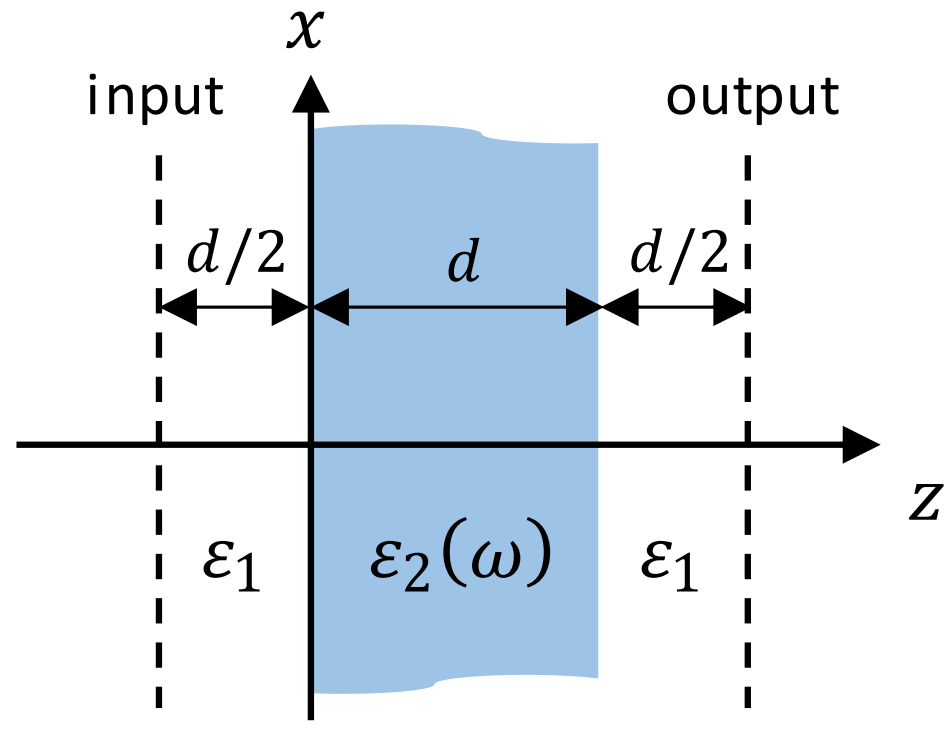

**Figure S1 Geometry and material parameters.** The superlens has a thickness denoted by $d$. Its relative permittivity is $\varepsilon_2(\omega)$, where $\omega$ is the angular frequency. The background relative permittivity, $\varepsilon_1$, is assumed to be independent of $\omega$. For compactness, we use the same notation for both relative permittivities and absolute permittivities, setting the vacuum permittivity to 1. All materials are non-magnetic, and we similarly set the vacuum permeability to 1. We use a Drude model for the superlens relative permittivity, $\varepsilon_2(\omega) = \varepsilon_\infty - \varepsilon_\infty \frac{\omega_p^2}{\omega^2 + i\omega\gamma}$, with $\varepsilon_\infty = \varepsilon_1 = 4$ , $\omega_p = 2.42 \times 10^{14}$ rad/s, and $\gamma = 0.056 \times 10^{14}$ rad/s. $\varepsilon_0$ and $\mu_0$ denote the vacuum permittivity and permeability.

## 2. QNM theory of superlens imaging

### 2.1 QNM electromagnetic fields

The superlens QNMs can be found via the vector Helmholtz equation under the constraint of tangential E- and H-field continuity at the interfaces $z = 0$ and $z = d$. Due to the translation-invariant symmetry, the QNM electromagnetic fields can be expressed as (with $\exp(-i\omega t)$ notation)

$$[\tilde{\mathbf{H}}(x,z,k_x) \quad \tilde{\mathbf{E}}(x,z,k_x)] = \exp[ik_x x] \begin{cases} [\tilde{\mathbf{h}}_{1,-} \quad \tilde{\mathbf{e}}_{1,-}] & (z < 0) \\ [\tilde{\mathbf{h}}_{2,-} + \tilde{\boldsymbol{h}}_{2,+} \quad \tilde{\mathbf{e}}_{2,-} + \tilde{\mathbf{e}}_{2,+}] & (0 \le z \le d), \\ [\tilde{\mathbf{h}}_{1,+} \quad \tilde{\mathbf{e}}_{1,+}] & (z > d) \end{cases} \tag{S2.1}$$

where $k_x$ is the parallel wavevector component and

$$\begin{bmatrix} \tilde{\mathbf{h}}_{1,-}(z,k_x) \\ \tilde{\mathbf{e}}_{1,-}(z,k_x) \end{bmatrix} = \tilde{h}_{1,-} \exp[-i\tilde{k}_{1z} z] \begin{cases} [0 \quad 1 \quad 0] \\ \frac{1}{\tilde{\omega}\varepsilon_1\varepsilon_0}[-\tilde{k}_{1z} \quad 0 \quad -k_x]' \end{cases} \tag{S2.2a}$$

$$\begin{bmatrix} \tilde{\mathbf{h}}_{2,-}(z,k_x) \\ \tilde{\mathbf{e}}_{2,-}(z,k_x) \end{bmatrix} = \tilde{h}_{2,-} \exp[-i\tilde{k}_{2z} z] \begin{cases} [0 \quad 1 \quad 0] \\ \frac{1}{\tilde{\omega}\varepsilon_2\varepsilon_0}[-\tilde{k}_{2z} \quad 0 \quad -k_x]' \end{cases} \tag{S2.2b}$$

$$\begin{bmatrix} \tilde{\mathbf{h}}_{2,+}(z,k_x) \\ \tilde{\mathbf{e}}_{2,+}(z,k_x) \end{bmatrix} = \tilde{h}_{2,+} \exp\left[i\tilde{k}_{2z}z\right] \begin{cases} [0 \quad 1 \quad 0] \\ \frac{1}{\tilde{\omega}\varepsilon_2\varepsilon_0}[\tilde{k}_{2z} \quad 0 \quad -k_x] \end{cases}, \tag{S2.2c}$$

$$\begin{bmatrix} \tilde{\mathbf{h}}_{1,+}(z,k_x) \\ \tilde{\mathbf{e}}_{1,+}(z,k_x) \end{bmatrix} = \tilde{h}_{1,+} \exp\left[i\tilde{k}_{1z}(z-d)\right] \begin{cases} [0 \quad 1 \quad 0] \\ \frac{1}{\tilde{\omega}\varepsilon_1\varepsilon_0}[\tilde{k}_{1z} \quad 0 \quad -k_x] \end{cases}. \tag{S2.2d}$$

The QNM-related quantities are denoted with a tilde. For instance, we denote by $\tilde{\omega}$ the complex-valued QNM eigenfrequency. Similarly, we have the following notations: $\tilde{k} = \tilde{\omega}/c$, $\tilde{k}_{1z} = \left(\tilde{k}^2 - k_x^2\right)^{1/2}$, $\tilde{k}_{2z} = \left(\tilde{k}^2 - k_x^2\right)^{1/2}$. Note that those expressions require a definition of a branch cut for the square root. We choose the sign of $\sqrt{x}$ such that $Re\left(\sqrt{x}\right) + Im\left(\sqrt{x}\right) > 0$.

The QNM amplitude coefficients are obtained by matching the tangential E- and H-field continuity conditions at the interfaces, $z = 0$ and $z = d$. For $H_y$, we obtain

$$\begin{aligned} \tilde{h}_{1,-} &= \tilde{h}_{2,+} + \tilde{h}_{2,-} \\ \tilde{h}_{1,+} &= \tilde{h}_{2,+}u + \tilde{h}_{2,-}u^{-1}, \end{aligned} \tag{S2.3}$$

with $\tilde{u} = \exp\left[i\tilde{k}_{2z}d\right]$. Furthermore, for $E_x$, we obtain

$$\begin{aligned} -\tilde{M}\tilde{h}_{1,-} &= \tilde{W}\tilde{h}_{2,+} - \tilde{W}\tilde{h}_{2,-} \\ \tilde{M}\tilde{h}_{1,+} &= \tilde{u}\tilde{W}\tilde{h}_{2,+} - \tilde{u}^{-1}\tilde{W}\tilde{h}_{2,-} \end{aligned}, \tag{S2.4}$$

with $\tilde{M} = \tilde{k}_{1z}/\varepsilon_1$ and $\tilde{W} = \tilde{k}_{2z}/\varepsilon_2(\tilde{\omega})$. In a matrix format, Eqs. (S2.3-S2.4) become

$$\begin{bmatrix} 1 & -1 & -1 & 0 \\ 0 & \tilde{u} & \tilde{u}^{-1} & -1 \\ -\tilde{M} & -\tilde{W} & \tilde{W} & 0 \\ 0 & \tilde{W}\tilde{u} & -\tilde{W}\tilde{u}^{-1} & -\tilde{M} \end{bmatrix} \begin{bmatrix} \tilde{h}_{1,-} \\ \tilde{h}_{2,+} \\ \tilde{h}_{2,-} \\ \tilde{h}_{1,+} \end{bmatrix} = 0. \tag{S2.5}$$

Because of symmetry with respect to the plane $z = d/2$, the modes are either symmetric or anti-symmetric:

$$\tilde{h}_{1,+} = \tilde{h}_{1,-},\ \tilde{h}_{2,+} = \frac{\tilde{h}_{1,-}}{\tilde{u}+1},\ \tilde{h}_{2,-} = \tilde{u}\frac{\tilde{h}_{1,-}}{\tilde{u}+1},\ -\frac{\tilde{M}}{\tilde{W}} = \frac{1-\tilde{u}}{1+\tilde{u}}, \tag{S2.6a}$$

for symmetric modes, and

$$\tilde{h}_{1,+} = -\tilde{h}_{1,-},\ \tilde{h}_{2,+} = -\frac{\tilde{h}_{1,-}}{\tilde{u}-1},\ \tilde{h}_{2,-} = \tilde{u}\frac{\tilde{h}_{1,-}}{\tilde{u}-1},\ -\frac{\tilde{M}}{\tilde{W}} = \frac{1+\tilde{u}}{1-\tilde{u}}, \tag{S2.6b}$$

for anti-symmetric modes.

## 2.2 The QNM dispersion relation

The superlens QNMs with complex-valued $\tilde{\omega}$ are found by looking for the non-trivial solutions of Eq. (S2.5) for which the matrix determinant is null. We obtain a transcendental equation, i.e. the dispersion relation $\tilde{\omega}(k_x)$,

$$\tilde{u} = \exp\left(i\tilde{k}_{2z}d\right) = \pm\frac{\varepsilon_2\tilde{k}_{1z}+\varepsilon_1\tilde{k}_{2z}}{\varepsilon_2\tilde{k}_{1z}-\varepsilon_1\tilde{k}_{2z}}. \tag{S2.7}$$

Note that the parameters, $\tilde{k}_{1z}$, $\tilde{k}_{2z}$, and $\varepsilon_2$, all depend on the frequency. The transcendental Eq. (S2.7), which includes an exponential term, admits an infinity of solutions with complex frequencies for every individual $k_x$.

To compute the surface polaritons of Fig. 2, we solve Eq. (S2.7) using an iterative procedure that is described in Appendix 2 in [3]. Convergence with an accuracy better than 1e-10 is achieved with typically 4-6 iterations starting from an initial guess value close to the pole.

Full analyticity can be restored, if necessary, by employing the closed-form approximation $\tilde{\omega}_\pm = \varpi\left(1 - \frac{\varepsilon_1}{8}\left(\frac{\omega_p}{ck_x}\right)^2\right) \pm \frac{\omega_p^2}{4\varpi}\exp(-k_x d)$, which is valid in the limit of large $k_x$.

## 2.3 QNM normalization

For reciprocal materials, as we consider here, the QNM are pairwise, implying that for every QNM with an in-plane wavevector component $k_x$, there exists another QNM with an opposite component $-k_x$. With the notation $\tilde{\mathbf{H}}(x,z,k_x) = \tilde{\mathbf{h}}(z,k_x)\exp(ik_x x)$ and $\tilde{\mathbf{E}}(x,z,k_x) = \tilde{\mathbf{e}}(z,k_x)\exp(ik_x x)$, we have [4]

$$\begin{cases} \tilde{\mathbf{h}}(z,k_x) = \tilde{\mathbf{h}}(z,-k_x) \\ \tilde{\mathbf{e}}(z,k_x)\cdot\hat{\mathbf{x}} = \tilde{\mathbf{e}}(z,-k_x)\cdot\hat{\mathbf{x}} \\ \tilde{\mathbf{e}}(z,k_x)\cdot\hat{\mathbf{z}} = -\tilde{\mathbf{e}}(z,-k_x)\cdot\hat{\mathbf{z}} \end{cases}. \tag{S2.8}$$

The QNM normalization factor $N$ (Eq. 2 in the main text) for a fixed in-plane wavevector component $k_x$ is given by [5]

$$N = \int_{-\infty}^{\infty}\left(\frac{\partial\omega\varepsilon}{\partial\omega}\varepsilon_0\tilde{\mathbf{e}}(z,k_x)\cdot\tilde{\mathbf{e}}(z,-k_x) - \mu_0\tilde{\mathbf{h}}(z,k_x)\cdot\tilde{\mathbf{h}}(z,-k_x)\right)dz. \tag{S2.9}$$

In general, the integral in Eq. (S2.9) is undefined because QNM fields exhibit exponential growth outside the superlens, necessitating some form of regularization—even in the case of one-dimensional slabs (see Annex 1 in [5]). For our specific problem, the surface polariton QNMs decay exponentially outside the superlens, allowing the integral to be evaluated either analytically or numerically without difficulty.

In the limit of large $|k_x|$, $k_{2z} \approx ik_x$ and $k_{0z} \approx ik_x$, and an analytical expression for $N$ can be derived:

$$N \approx -\frac{\mu_0}{k_x}\frac{\varepsilon_1-\varepsilon_2}{\varepsilon_1}\left(2 + k_x d\frac{\varepsilon_1+\varepsilon_2}{\varepsilon_1}\right)\tilde{h}_{1,-}^2 + \frac{k_x}{\varepsilon_0}\frac{2}{\varepsilon_2\varepsilon_1}\frac{(\varepsilon_2-\varepsilon_\infty)^2}{\omega_p^2\varepsilon_\infty}\left(2 + i\frac{\gamma}{\tilde{\omega}}\right)\tilde{h}_{1,-}^2. \tag{S2.10}$$

By Further neglecting small terms under the assumptions $\varepsilon_1 = \varepsilon_\infty$, $|\tilde{\omega}| \gg \gamma$, $(\varepsilon_1 + \varepsilon_2(\tilde{\omega})) \to 0$, and $|\omega_p/c| \ll |k_x|$, we can derive at an even simpler expression:

$$N \approx -\frac{16k_x}{\varepsilon_0\omega_p^2\varepsilon_1}\tilde{h}_{1,-}^2. \tag{S2.11}$$

## 2.4 The QNM excitation coefficient

The QNM excitation coefficients are given by (Eq. (4) in the main text)

$$\alpha_m(\omega,k_x) = \varepsilon_0\left[\frac{\tilde{\omega}_m}{\tilde{\omega}_m-\omega}(\varepsilon_2(\tilde{\omega}_m)-\varepsilon_1) + (\varepsilon_1-\varepsilon_\infty)\right]\int_0^d\tilde{\mathbf{E}}_m(z,x,-k_x)\cdot\mathbf{E}_{inc}\,dz, \tag{S2.13}$$

where $\mathbf{E}_{inc}$ is the electric field of a p-polarized plane wave with a frequency $\omega$ and an in-plane wave vector $k_x$, as defined in Eq. (S3.1). $m = s\ or\ a$ represents symmetric or anti-symmetric QNMs.

The integral in the lens can be computed analytically. We find

$$\int_0^d\tilde{\mathbf{E}}_m(z,x,-k_x)\cdot\mathbf{E}_{inc}\,dz = Z_0\varepsilon_1^{-1/2}H_{inc}\frac{[\exp[i(\tilde{k}_{2mz}+\tilde{k}_{1mz})d]-1]}{i(\tilde{k}_{2mz}+\tilde{k}_{1mz})\tilde{\omega}_m\varepsilon_2(\tilde{\omega}_m)\varepsilon_0\tilde{k}_{1m}}\left(-k_x^2+\tilde{k}_{1mz}\tilde{k}_{2mz}\right)\begin{cases}\frac{\tilde{h}_{1,-,m}}{\tilde{u}_m+1}\\ \frac{\tilde{h}_{1,-,m}}{1-\tilde{u}_m}\end{cases}$$

$$+Z_0\varepsilon_1^{-1/2}H_{inc}\frac{[\exp[i(\tilde{k}_{1mz}-\tilde{k}_{2mz})d]-1]}{i(\tilde{k}_{1mz}-\tilde{k}_{2mz})\tilde{\omega}_m\varepsilon_2(\tilde{\omega}_m)\varepsilon_0\tilde{k}_{1m}}\left(-k_x^2-\tilde{k}_{1mz}\tilde{k}_{2mz}\right)\begin{cases}\frac{\tilde{u}_m\tilde{h}_{1,-,m}}{\tilde{u}_m+1}\ (m=s)\\ \frac{\tilde{u}_m\tilde{h}_{1,-,m}}{\tilde{u}_m-1}\ (m=a)\end{cases}, \tag{S2.14}$$

where $Z_1 = \sqrt{\mu_0/\varepsilon_0}$ denotes the impedance of the background medium. The expression for $N$ takes the upper and lower terms for symmetric and anti-symmetric QNMs, respectively.

An approximate and simple formula for $\alpha_m$ is given by

$$\alpha_m(\omega) = \left[\frac{-2k_x}{\tilde{\omega}_m-\omega}\right]\frac{H_{inc}}{\omega\varepsilon_0\varepsilon_1}, \tag{S2.15}$$

under the following conditions: $|\tilde{k}d| \ll 1$, $|k_x d| \gg 1$, and $(\varepsilon_1 + \varepsilon_2(\tilde{\omega}_m)) \to 0$. $H_{inc}$ will be defined in section 3.

### 2.5 Numerical test of the accuracy of Eq. (5)

In this subsection, we evaluate the accuracy of Eq. (5). We consider three test cases inspired by recent experimental results in the thermal infrared regime [6]. The green curve in Fig. S2 is obtained for a real frequency illumination, $\lambda = 11.5$ µm, the blue curve for the same real frequency assuming a lossless superlens (i.e., $\gamma = 0$), and the dotted red curve for a complex frequency illumination $\omega = 2\pi c/\lambda - i\gamma/2$, such that the imaginary part of $\varepsilon_2(\omega)$ vanishes [6]. All three curves are computed using Eq. (5) and compared against reference data computed using a $2 \times 2$ matrix formalism of Suppl. Note 3. These reference data are shown in insets A, B, and C using circles, squares, and crosses.

The comparison confirms our expectations. At small $k_x$'s (inset A), the model shows noticeable discrepancies, though it qualitatively captures key features such as the peak around $k_x/q \approx 1.2$. For intermediate $k_x$'s (inset B), the relative error is small, on the order of 1%, while at large $k_x$'s, the agreement becomes excellent. Additional results that further demonstrate the high accuracy of Eq. (5) at large $k_x$ are provided in Fig. S4.

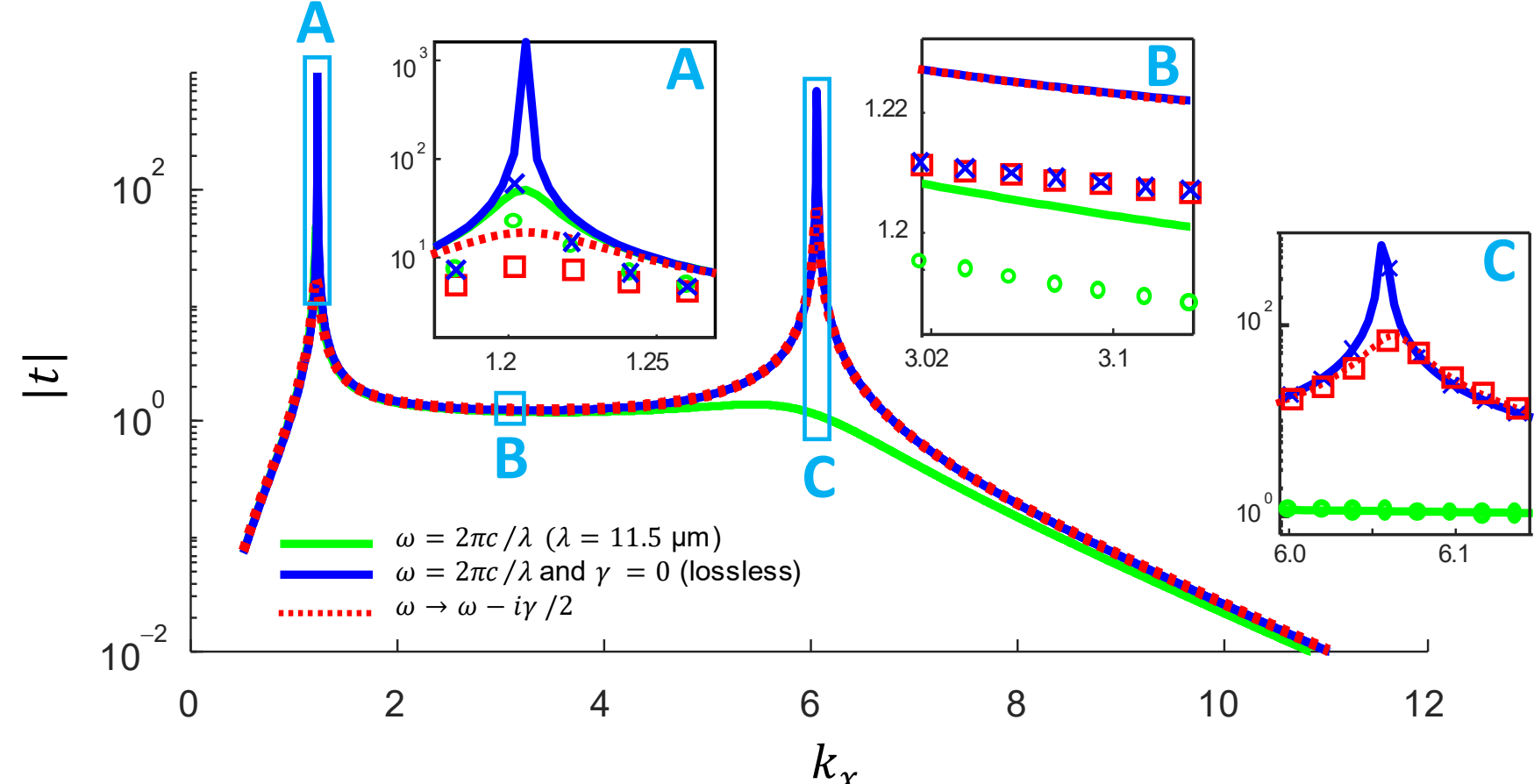

**Figure S2 Test of the accuracy of Eq. (5).** $|t|$ predicted with the analytical model, Eq. (5), for an incident wavelength of $\lambda = 11.5$ µm (green), for the same wavelength assuming that the Drude superlens is lossless ($\gamma = 0$) (blue), and for a complex frequency illumination $\omega = 2\pi c\lambda^{-1} - i\gamma/2$ (dashed red), such that the imaginary part of $\varepsilon_2(\omega)$ is null. Insets: enlarged views of 3 zones. Cross, circles and squares are reference data obtained with the $2 \times 2$ -transfer matrix formalism.

The results in Fig. S2, which are obtained for a frequency slightly offseted compared to the lens resonance, are simply explained. The two prominent peaks of the transfer function arise from the shape of the polaritonic dispersion curves in Fig. 1, which intersects the horizontal line for $\lambda = 11.5$ µm at two distinct $k_x$ values, first at small $k_x$ with the symmetric polariton $\tilde{\omega}_+$ and then at $k_x \approx 6q$ with the antisymmetric one.

## 3. Direct computation of the superlens response with 2×2 transfer matrices

This Section contains classical results. The superlens is assumed to be illuminated by a p-polarized plane wave with a frequency $\omega$ and an in-plane wave vector $k_x$

$$\begin{bmatrix}\mathbf{H}_{inc}(x,z,k_x)\\ \mathbf{E}_{inc}(x,z,k_x)\end{bmatrix} = H_{inc}\exp[i(k_{1z}z + k_x x)]\begin{cases}[0 \quad 1 \quad 0]^T\\ \frac{1}{\omega\varepsilon_1\varepsilon_0}[k_{1z} \quad 0 \quad -k_x]^T\end{cases}, \tag{S3.1}$$

We use the following notations: $k = \omega/c$, $k_{1z} = (\varepsilon_1 k^2 - k_x^2)^{1/2}$, $k_{2z} = (\varepsilon_2(\omega)k^2 - k_x^2)^{1/2}$ and $[\mathbf{h}_{inc}, \mathbf{e}_{inc}] = \exp(-ik_x x)\,[\mathbf{H}_{inc}, \mathbf{E}_{inc}]$. With the same notations, the reflected and transmitted plane waves are respectively denoted

$$\begin{bmatrix}\mathbf{h}_r(z,k_x)\\ \mathbf{e}_r(z,k_x)\end{bmatrix} = rH_{inc}\exp(-ik_{1z}z)\begin{cases}[0 \quad 1 \quad 0]^T\\ \frac{1}{\omega\varepsilon_1\varepsilon_0}[-k_{1z} \quad 0 \quad -k_x]^T\end{cases}, \tag{S3.2a}$$

$$\begin{bmatrix} \mathbf{h}_t(z,k_x) \\ \mathbf{e}_t(z,k_x) \end{bmatrix} = tH_{inc}\exp(ik_{1z}z)\begin{cases} [0 \quad 1 \quad 0]^T \\ \frac{1}{\omega\varepsilon_1\varepsilon_0}[k_{1z} \quad 0 \quad -k_x]^T, \end{cases} \tag{S3.2b}$$

The field in the superlens is a superposition of two plane waves

$$\begin{bmatrix} \mathbf{h}_{2,-}(z,k_x) \\ \mathbf{e}_{2,-}(z,k_x) \end{bmatrix} = a_{2,-}\, H_{inc}\exp(-ik_{2z}z)\begin{cases} [0 \quad 1 \quad 0] \\ \frac{1}{\omega\varepsilon_2\varepsilon_0}[-k_{2z} \quad 0 \quad -k_x]' \end{cases} \tag{S3.2c}$$

$$\begin{bmatrix} \mathbf{h}_{2,+}(z,k_x) \\ \mathbf{e}_{2,+}(z,k_x) \end{bmatrix} = a_{2,+}\, H_{inc}\exp(+ik_{2z}z)\begin{cases} [0 \quad 1 \quad 0] \\ \frac{1}{\omega\varepsilon_2\varepsilon_0}[k_{2z} \quad 0 \quad -k_x]' \end{cases} \tag{S3.2d}$$

The superlens response can be directly obtained by writing the tangential field continuities at $z = 0$ and $z = d$. At $z = 0$, we have $1 + r = a_{2-} + a_{2+}$ and $M(1 - r) = W(-a_{2,-} + a_{2,+})$. At $z = d$, we have $tu_1 = u_2^{-1}a_{2-} + u_2 a_{2+}$ and $tMu_1 = -Wu_2^{-1}a_{2-} + Wu_2 a_{2+}$. In a matrix format,

$$\begin{bmatrix} 1 & -1 & -1 & 0 \\ 0 & u_2^{-1} & u_2 & -u_1 \\ M & -W & W & 0 \\ 0 & Wu_2^{-1} & -Wu_2 & Mu_1 \end{bmatrix}\begin{bmatrix} r \\ a_{2-} \\ a_{2+} \\ t \end{bmatrix} = \begin{bmatrix} -1 \\ 0 \\ M \\ 0 \end{bmatrix}. \tag{S3.3}$$

with $u_2 = \exp(ik_{2z}d)$, $u_1 = \exp(ik_{1z}d)$, $M = k_{1z}/\varepsilon_1$ and $W = k_{2z}/\varepsilon_2$.

Solving this equation, one readily finds that the coefficient for $t$ and $r$

$$r = -1 + \frac{2M(M+W) - 2M(M-W)u_2^2}{(M+W)^2 - (M-W)^2 u_2^2}, \tag{S3.4a}$$

$$t = u_1^{-1}\frac{-4MW}{(M-W)^2 u_2 - (M+W)^2 u_2^{-1}}. \tag{S3.4b}$$

In terms of transfer function from plane $z = -d/2$ to $z = 3d/2$, the coefficients $t$ and $r$ become $r \equiv u_1 r$ and $t \equiv u_1^2 t$. We finally obtain:

$$\frac{h_r(z=-d/2,k_x)}{h_{inc}(z=-d/2,k_x)} = r\ (\equiv u_1 r) = -u_1 + u_1\frac{2M(M+W) - 2M(M-W)u_2^2}{(M+W)^2 - (M-W)^2 u_2^2}, \tag{S3.5a}$$

$$\frac{h_t(z=3d/2,k_x)}{h_{inc}(z=-d/2,k_x)} = t\ (\equiv u_1^2 t) = \frac{-4MW}{(M-W)^2 u_2 - (M+W)^2 u_2^{-1}}u_1. \tag{S3.5b}$$

These formulae have been verified numerically by comparison with the matrix inversion of Eq. (S3.3). However, since $u_2$ and $u_2^{-1}$ are respectively very small and large for large $k_x$'s, the matrix in Eq. (S3.3) is close to singular or badly scaled, and therefore we use the freeware RETICOLOfilm-stack [7] that relies on enhanced S-matrix products that are unconditionally stable. The freeware is written in Matlab. The RETICOLO program used to test the approximate model and compute the reference data in Fig. S3 is simply:

```
% computation of reference data with RETICOLOfilm stack
clear
wavelength=11.5+i*0.5; % in µm
thickness=0.440; % superlens thickness
pol=-1; % TM polarization (TE: pol=1)
k0=2*pi/wavelength;
nSiC = lalindice(ldinc);
nSiO2 = 2;
beta0 = linspace(0,20,1000); % k//
res0_0D(pol, k0, beta0); % the Matlab program is vectorialized
a_SiO2=res1_0D(nSiO2); a_SiC=res1_0D(nSiC);
s_SiO2 = res2_0D(a_SiO2,thickness/2); % S-matrix of the silica layer
s_SiC = res2_0D(a_SiC,thickness); % S-matrix of the silica layer
[result, su, sb]=res2_0D(s_SiO2*s_SiC*s_SiO2, a_SiO2, a_SiO2);
t = result.inc_top_transmitted.amplitude; % t in Eq. (S3.5a)
r = result.inc_top_reflected.amplitude;   % r in Eq. (S3.5b)
figure; plot(beta0,abs(t),'b','linewidth',2); xlabel('k_{||} [1/µm]'); ylabel('|t|');

% Drude model for the superlens
function index=lalindice(wavelength)
```

```
epsinf=4;omegaP=2.4217*10^14/3e8;gamma=0.0565*10^14/3e8;
k=2*pi./(wavelength*1e-6);
eps2=epsinf*(1-omegaP^2./(k.^2+i*k*gamma));
index=sqrt(eps2);
end
```

Figure S3 compares the closed-form expression for the transmission in Eq. (S3.5b) with numerical data obtained with the program for an illumination frequency equal to the polariton frequency $\varpi$ of a single interface $\varepsilon_2/\varepsilon_1$: $\varepsilon_2(\varpi) = -\varepsilon_1$. Perfect agreement with numerical precision is achieved.

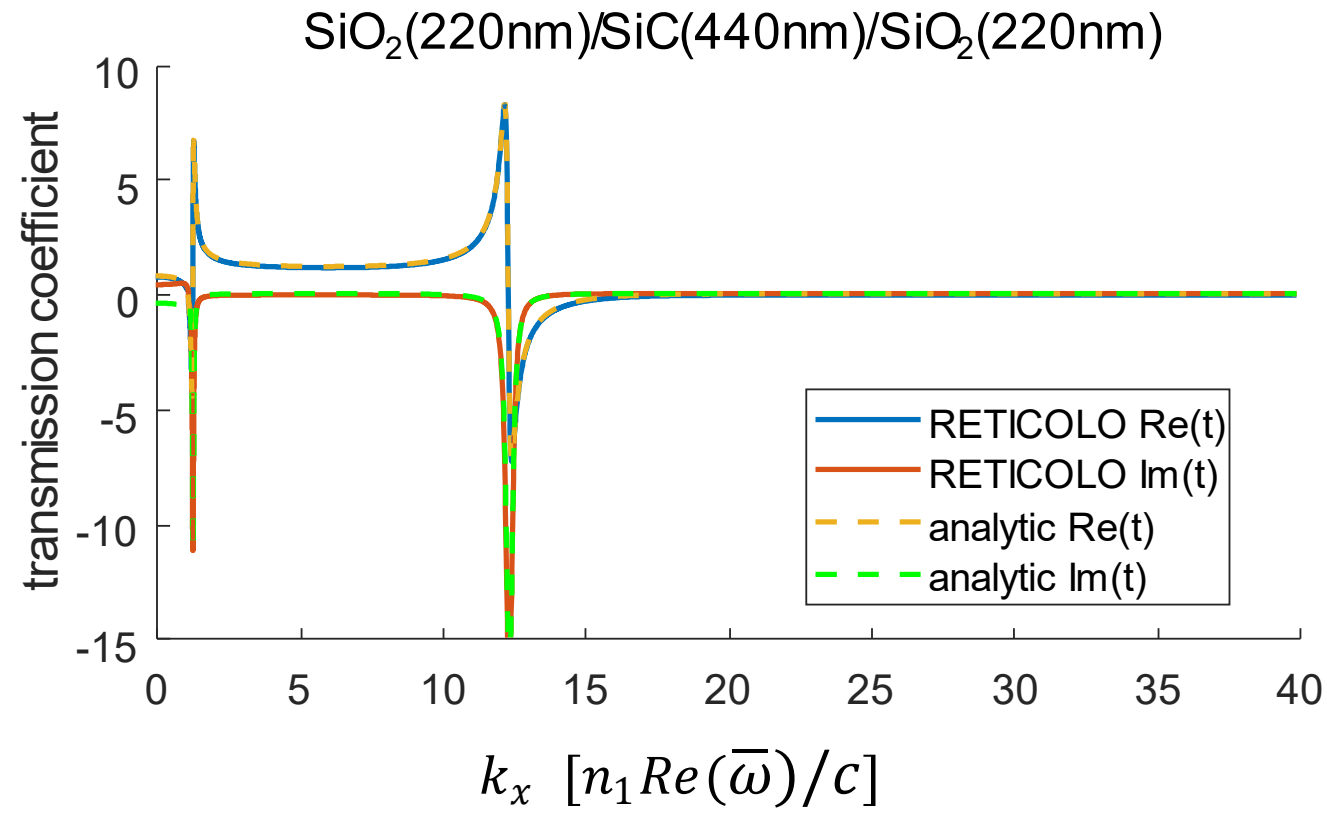

**Figure S3.** Comparing the closed-form expression for the transmission in Eq. (S3.5b) with numerical data obtained with the freeware RETICOLOfilm-stack [7]. The comparison is performed for an incident plane wave with a complex-valued frequency $\varpi$ such that $\varepsilon_2(\varpi) = -\varepsilon_1$, i.e. $\bar{\lambda} \approx 11.006 + i0.1816$.

## 4. Questioning loss compensation with complex-frequency illumination

By transforming the frequency into a suitable complex value $\omega \to \omega - i\gamma/2$, the permittivity can be turned into a purely real value $\varepsilon(\omega) = 1 - \omega_p^2/\left(\omega^2 + \frac{\gamma^2}{4}\right)$. This mathematical result forms the basis for the widely held view [6,8-9] that complex frequencies with $\mathrm{Im}(\omega) = \gamma/2$ effectively mimics loss removal.

However, the transformation $\omega \to \omega - i\gamma/2$ only approximately eliminates absorption losses. In reality, it creates an artificial medium that exhibits amplification, as both its effective permittivity and permeability become complex, with $Im(\tilde{\omega}\varepsilon) < 0$ and $Im(\tilde{\omega}\mu) < 0$.

To illustrate this, let us consider a general scattering problem where a scatterer is illuminated by a source $\mathbf{J}(\boldsymbol{r})$ emitting at a real frequency $\omega$ for simplicity. The scatterer is assumed to have a complex permittivity $\boldsymbol{\varepsilon}(\boldsymbol{r},\omega)$ and permeability $\mu(\boldsymbol{r},\omega)$ because it is lossy. The Maxwell's equations read as

$$\nabla \times \mathbf{E}(\boldsymbol{r}) = i\omega\mu(\boldsymbol{r},\omega)\mu_0\mathbf{H}(\boldsymbol{r}) \tag{S4.1a}$$

$$\nabla \times \mathbf{H}(\boldsymbol{r}) = -i\omega\varepsilon(\boldsymbol{r},\omega)\varepsilon_0\mathbf{E}(\boldsymbol{r}) + \boldsymbol{J}(\boldsymbol{r}). \tag{S4.1b}$$

Let us imagine that we can find a complex frequency $\tilde{\omega}$ such that $\boldsymbol{\varepsilon}(\boldsymbol{r},\tilde{\omega})$ is real. For a Drude scatterer, we simply have $\tilde{\omega} = \omega - i\gamma/2$. At this complex frequency, the Maxwell equations become

$$\nabla \times \mathbf{E}(\boldsymbol{r}) = i\tilde{\omega}\mu(\boldsymbol{r},\tilde{\omega})\mu_0\mathbf{H}(\boldsymbol{r}) \tag{S4.2a}$$

$$\nabla \times \mathbf{H}(\boldsymbol{r}) = -i\tilde{\omega}\varepsilon(\boldsymbol{r},\tilde{\omega})\varepsilon_0\mathbf{E}(\boldsymbol{r}) + \boldsymbol{J}(\boldsymbol{r}). \tag{S4.2b}$$

Suppose that $\mu(\boldsymbol{r},\tilde{\omega}) = 1$ for simplicity. In this case, an interpretation of Eqs. (S4.2a-b) is that the complex frequency field within the Drude scatterer interacts with a non-magnetic medium that has a real-valued permittivity. While this interpretation is correct, it does not imply that absorption losses have been eliminated. As we will demonstrate by applying the Poynting theorem, loss cancellation has not, in fact, occurred.

Let us now consider a second set of Maxwell's equations, obtained by complex conjugating the previous equations,

$$\nabla\times\mathbf{E}^* = i(-\tilde{\omega}^*)\mu(\boldsymbol{r},-\tilde{\omega}^*)\mu_0\mathbf{H}^*, \tag{S4.3a}$$

$$\nabla\times\mathbf{H}^* = -i(-\tilde{\omega}^*)\varepsilon(\boldsymbol{r},-\tilde{\omega}^*)\varepsilon_0\mathbf{E}^* + \boldsymbol{J}^*(\boldsymbol{r}), \tag{S4.3b}$$

obtained by using $\varepsilon^*(\boldsymbol{r},\omega) = \varepsilon(\boldsymbol{r},-\omega^*)$ and $\mu^*(\boldsymbol{r},\omega) = \mu(\boldsymbol{r},-\omega^*)$ due to the Hermitian symmetry of real Fourier transforms. Note that the Poynting vectors associated to Eqs. (S4.2a-b) and (S4.3a-b) are identical, implying that these equations also satisfy the same outgoing wave condition.

We now apply the Lorentz reciprocity theorem to the vector $\mathbf{E}^*\times\mathbf{H} + \mathbf{E}\times\mathbf{H}^*$ over a volume $\Omega$, assuming that the source $\mathbf{J}$ (or $\mathbf{J}^*$) lies outside this volume, and obtain:

$$\frac{1}{2}\mathrm{Re}\iint_{\Sigma}(\mathbf{E}\times\mathbf{H}^*)\cdot\mathbf{n}\,dS = -\frac{1}{4}\iiint_{\Omega}\left(\mathrm{Im}(\tilde{\omega}\varepsilon)\varepsilon_0|\mathbf{E}|^2 + \mathrm{Im}(\tilde{\omega})\mu_0|\mathbf{H}|^2\right)d\Omega. \tag{S4.4}$$

For the usual case of real frequencies, $\tilde{\omega}$ can be factored out of the imaginary parts and shows up as a prefactor to the integral $(1/4 \rightarrow \tilde{\omega}/4)$.

The left-hand side of Eq. (S4.4) represents the real part of the surface integral of the Poynting vector, corresponding to the power flowing out through the surface $\Sigma$. In the case of a non-absorbing material enclosed within $\Sigma$, the net power flow across the boundary must be zero—implying that inflow is exactly balanced by outflow. This leads to the condition:

$$\mathrm{Im}(\tilde{\omega}\varepsilon) = \mathrm{Im}(\tilde{\omega}\mu) = 0. \tag{S4.5}$$

If the material within $\Sigma$ is absorbing, the right-hand term in Eq. (S4.4) is negative, in agreement with the interpretation that more power enters the volume than exits it. Thus, $\mathrm{Im}(\varepsilon) > 0$ and Eq. (S4.4) expresses that the flux of the Poynting vector through the surface (left-hand side) equals the Ohmic losses inside the volume (right-hand side). This interpretation assumes the time dependence $\exp(-i\omega t)$, under which an amplifying medium corresponds to $\mathrm{Im}(\varepsilon) < 0$.

By choosing a complex frequency such that $\varepsilon(\tilde{\omega})$ is real, one implements $Im(\tilde{\omega}\varepsilon) < 0$ and $Im(\tilde{\omega}\mu) < 0$, indicating that the Poynting flux is positive—i.e., the medium amplifies through both its permittivity and permeability.

The conditions (S4.5) have a clear interpretation when the integration contour $\Sigma$ delimitates a homogeneous medium: the complex propagation constant $\tilde{k} = \sqrt{\tilde{\omega}^2\varepsilon\mu\varepsilon_0\mu_0}$ becomes real-valued. Loss compensation does not require $\varepsilon(\tilde{\omega})$ to be real, but rather that both $\tilde{\omega}\varepsilon$ and $\tilde{\omega}\mu$ be real.

A further complication arises when the integration contour $\Sigma$ encompasses multiple materials. In such cases, the conditions (S4.5) are harder to satisfy, since $\varepsilon$ and $\mu$ generally vary with position.

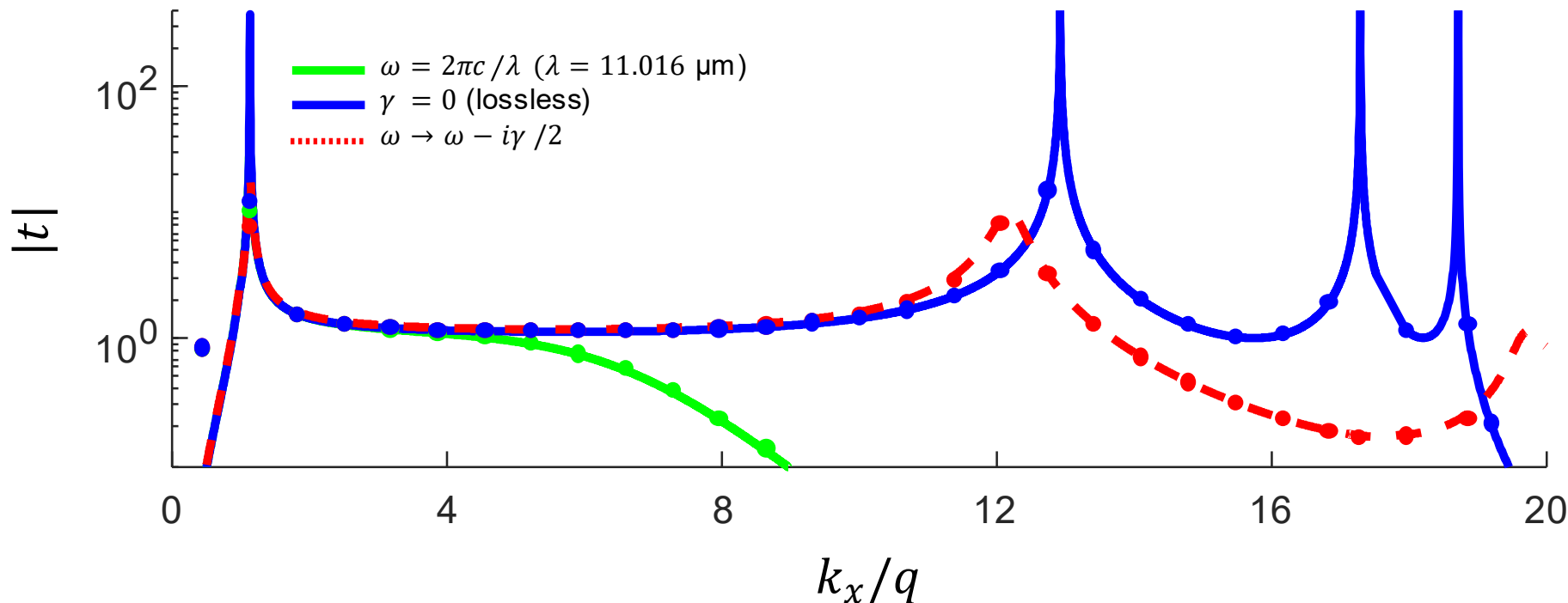

**Figure S4.** The significant discrepancy between the blue ($\gamma = 0$) and dotted red ($\omega \rightarrow \omega - i\gamma/2$) curves underscores that loss compensation cannot be rigorously achieved using complex frequency illuminations, as predicted by our analysis using the Poynting theorem. Note that all curves are obtained with the QNM model, Eq. (5). They are superimposed with the curve for $k_x > 1.5q$ with reference data (circles) obtained with the $2\times2$ matrix formalism.

In Fig. S2, minor discrepancies were observed at the resonance peaks between the blue and dotted red curves, respectively obtained for a real frequency illumination, $\lambda = 11.5$ μm and $\gamma = 0$, and a complex frequency illumination $\omega = 2\pi c/\lambda - i\gamma/2$. This relative agreement was fortuitous, as shown in Fig. S4, which is obtained for a slightly different wavelength, $\lambda = 11.0160$ μm, corresponding to $\omega = \mathrm{Re}[\tilde{\omega}_+(20q)]$. It now highlights a significant difference between the blue and dotted red curves, especially for large spatial frequencies, $k_x > 11q$. Compensation of absorption loss cannot be implemented with complex frequency illuminations, especially in resonant systems.

## 5. Observability of steady state regimes for complex frequency illuminations

This section investigates the observability of steady-state regimes for complex frequency illuminations. We consider the general case of a resonator with a known spectrum $\{\tilde{\omega}_1, \tilde{\omega}_2 \ldots\}$, without imposing any assumptions on the QNM fields or the geometry of the system. The resonator is positioned around its 'center of mass' $\boldsymbol{r_c}$. We will take $\boldsymbol{r_c} = \boldsymbol{0}$, without loss of generality. We also assume that all the resonator materials are non-dispersive.

Direct illuminations at complex frequencies (ω) can only be applied over a semi-infinite time interval. Consequently, the incident wavepacket electric field typically takes the form: $\boldsymbol{E}_{inc}(\boldsymbol{r},t) = \mathbf{E}_0\, S\left(t - \boldsymbol{r}\cdot\frac{\boldsymbol{u}}{c}\right)\exp\left[-i\omega\left(t - \boldsymbol{r}\cdot\frac{\boldsymbol{u}}{c}\right)\right]$, where $\boldsymbol{u}$ is a unit vector in the direction of the incident wavevector, and $S(t)$ a sigmoid function, such that the wavepacket reaches the resonator positioned at the origin at $t = 0$.

We begin by recalling recent results on QNM expansions in the time domain [10,11]. When the resonator's frequency-domain response is known via a QNM expansion—see Eq. (3) in the main text—the response to any incident wavepacket can be determined by decomposing the pulse spectrally and summing over all contributing frequencies. Consequently, the electric field $\mathbf{E}_S(\boldsymbol{r},t)$ in response to the wavepacket can be expressed as a sum of QNM contributions [10]

$$\mathbf{E}_S(\boldsymbol{r},t) = Re\left(\textstyle\sum_m \beta_m(t)\tilde{\mathbf{E}}_m(\boldsymbol{r})\right). \tag{S5.1}$$

The validity of the expansion in Eq. (S5.1) has been rigorously established, even for complex geometrical configurations including dispersive resonators. For non-dispersive resonator, it has been recently shown [10,11] that the time-dependent modal coefficients are given by

$$\beta_m(t) = i\tilde{\omega}_m \int_{-\infty}^{t} O_m(t')\exp\left(i\tilde{\omega}_m(t'-t)\right)dt' - O_m(t), \tag{S5.2}$$

where the overlap function $O_m(t) = \varepsilon_0\langle \boldsymbol{E}_{inc}(\boldsymbol{r},t)|\Delta\varepsilon(\boldsymbol{r})|\tilde{\boldsymbol{E}}_m(\boldsymbol{r})\rangle$ quantifies the interaction between the incident wavepacket and the QNM field, integrated over the resonator volume $V_{res}$ where the permittivity change is $\Delta\varepsilon(\boldsymbol{r})$.

To simplify the treatment and achieve full analyticity, we assume that the incident wavepacket varies slowly over the spatial extent of the resonator and the envelope $S(t)$ is a slowly varying sigmoid function. The implications of these two assumptions on the resonator response are discussed in detail in [10], where it is shown that the generality or validity of our conclusions are not compromised. Under this approximation, recalling that the center of mass of the resonator is at the origin, the overlap integral $\langle \boldsymbol{E}_{inc}|\Delta\boldsymbol{\varepsilon}(\boldsymbol{r})|\tilde{\mathbf{E}}_m\rangle$ simplifies to $\langle \mathbf{E}_0|\Delta\boldsymbol{\varepsilon}|\tilde{\mathbf{E}}_m\rangle\, S(t)\exp[-i\omega t]$. Elementary algebra then leads to

$$\beta_m(t) = \varepsilon_0\langle \mathbf{E}_0|\Delta\boldsymbol{\varepsilon}|\tilde{\mathbf{E}}_m\rangle\, S(t)\left\{\frac{\omega}{\tilde{\omega}_m-\omega}\exp(-i\omega t) - \frac{\tilde{\omega}_m}{\tilde{\omega}_m-\omega}\exp(-i\tilde{\omega}_m t)\right\} \tag{S5.3}$$

for $\omega \neq \tilde{\omega}_m$.

In Eq. (S5.3), the first exponential term inside the brackets represents the *transient response*, which decays over time at a rate set by the imaginary part of the QNM frequency. The second term corresponds to the *steady-state regime*, characterized by sustained oscillations at the frequency of the driving wavepacket.

In the case where the illumination frequency is close to the fundamental resonance frequency with the longest lifetime ($\omega \to \tilde{\omega}_m$), the expression for $\beta_m$ becomes

$$\beta_m(t) = i\omega t\varepsilon_0\, S(t)\, \langle \mathbf{E}_0|\Delta\boldsymbol{\varepsilon}|\tilde{\mathbf{E}}_m\rangle\exp(-i\omega t), \tag{S5.4}$$

implying that the global modal response (the sum of the study-state and transient contributions) no longer diverges as $(\tilde{\omega}_m - \omega)^{-1}$. Its linear growth in time reflects a resonant divergence, indicating that the system accumulates energy indefinitely in the absence of saturation mechanisms.

Together, Eqs. (S5.1) and (S5.3) yield closed-form expressions for the time-domain scattered field $\mathbf{E}_S(\boldsymbol{r},t)$, enabling a direct analysis of the conditions under which steady-state regimes can be observed for complex-frequency excitations.

### 5.1 General discussion

To investigate the conditions under which quasi-steady-state regimes are observable, we examine a representative example where the resonator's spectrum consists of three dominant QNMs, with complex eigenfrequencies, $\tilde{\omega}_m$ with $m = 1,2,3$. These modes are represented by blue circles in the lower half of the complex frequency plane shown in Fig. S5.1a. We consider three complex-frequency illuminations, indicated by red crosses and labeled (1), (2), and (3). For each case, the scattered field $\mathbf{E}_S(\boldsymbol{r}, t)$ is computed using Eq. (S5.1), incorporating all three QNM contributions to capture both transient and steady-state behaviors.

In Fig. S5.1b, the transient components are represented by blue curves. They consistently decay in time. The steady-state components, depicted in green, oscillate at the excitation frequency and display distinct long-time behaviors—growing in case (1), and decaying in cases (2) and (3).

We start our analysis with a complex frequency excitation characterized by a positive imaginary part (labelled as (1) in Fig. S5.1a). Such excitations are particularly interesting because they can produce strong scattering responses not accessible through real-frequency illumination—for instance, by interacting with complex scattering zeros located in the *upper half* of the complex frequency plane [8,9]. While the transient component decays over time, the quasi-steady-state response grows exponentially, guaranteeing its observability in principle, even at "any" times.

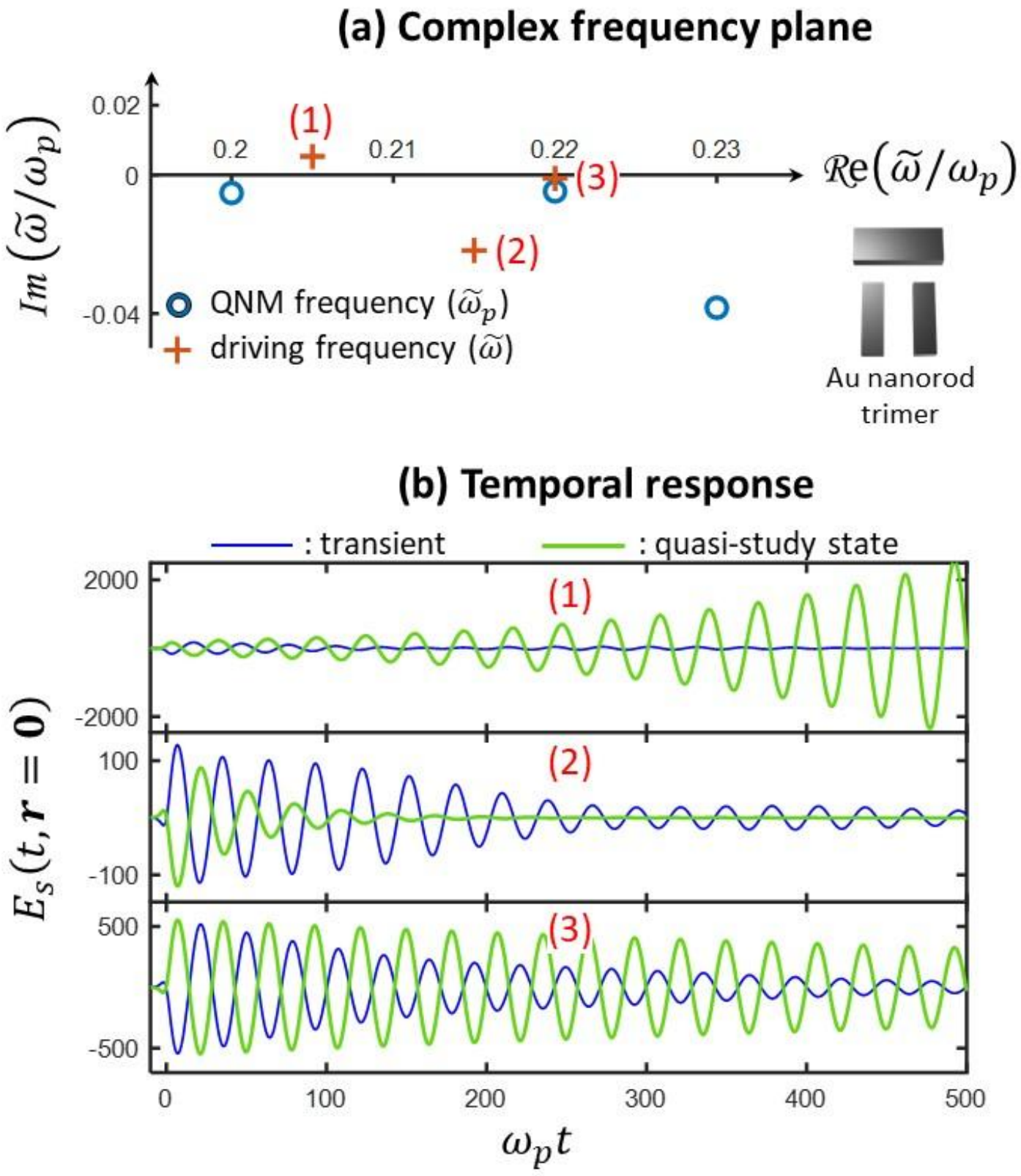

**Figure S5. Observability of quasi-study state regimes under complex-frequency excitations.** **(a)** Complex frequency plane with three dominant poles (blue circles) in the visible frequency range. **(b)** Temporal evolution of the field scattered by the resonator for three complex frequency excitations, indicated by red plusses and labelled as (1), (2), and (3). The blue curves represent the transients, while the green curves illustrate the quasi-steady state regimes. Note that these plots serve as mere illustrations, as the exact resonator response depends on the QNM fields at the observation point $\boldsymbol{r}$ (Eq. (6.1)) and the overlap integral between the incident wavepacket electric field and the normalized QNM, which are all fixed to a constant value independent of the QNM. The three poles are representative of the dominant QNMs of a nanoparticle (left inset) composed of three gold nanorods. Specific values are : $\omega_1/\omega_p = 0.20 \times (1 - i/(2Q_1))$ rad/s, $\omega_2/\omega_p = 0.22 \times (1 - i/(2Q_2))$ rad/s and $\omega_3/\omega_p = 0.23 \times (1 - i/(2Q_3))$ rad/s with $Q_1 = 19$, $Q_2 = 23$ and $Q_3 = 9$, $\omega_p = 1.4\times10^{16}$ rad s$^{-1}$. More details are found in Fig. 7 in [4].

In the second scenario, case (2), the excitation has a negative imaginary part, indicating temporal decay. If its damping rate $\mathrm{Im}(\omega)$ exceeds that of the fundamental mode (the QNM with the longest lifetime), the steady-state regime becomes unobservable, as it decays faster than the transient of the fundamental mode. However, exceptions may arise—for example, when the fundamental mode is not excited due to symmetry constraints or selection rules that suppress its coupling to the incident field.

Our final scenario, case (3), is of particular significance for superlens imaging—specifically, when the complex excitation frequency lies close to the eigenfrequency of the fundamental mode, $\omega \approx \tilde{\omega}_2$. Due to the minus sign in the bracketed term of Eq. (S5.3) reveals that the transient response can substantially obscure the steady-state regime. This is illustrated in the lower panel of Fig. S5.1b, where the steady-state behavior becomes discernible only after a prolonged duration when $Im(\omega)$ slightly exceeding $Im(\tilde{\omega}_2)$). Consequently, the emergence of a steady-state regime becomes apparent only after an extended duration with a weak signal susceptible to noise contamination, posing challenges for practical detection.

## 5.2 Specific case of the superlens

The superlens resonances with longer lifetimes correspond to smaller $k_x$ values (Fig. S6a). This suggests that the steady-state regime at low $k_x$ may not be observable due to the dominance of the transient regime. Figure S6 demonstrates this behavior for our optimal frequency $\tilde{\omega}_+(17q)$ (blue curve). Only the spatial frequency ranges where the steady-state regime persists longer than the transient regime are shown.

As the imaginary parts of the complex-frequency illuminations are progressively reduced, the spatial frequency range increases. However, this comes at the cost of a decreasing magnitude of the transmission coefficient $|t|$.

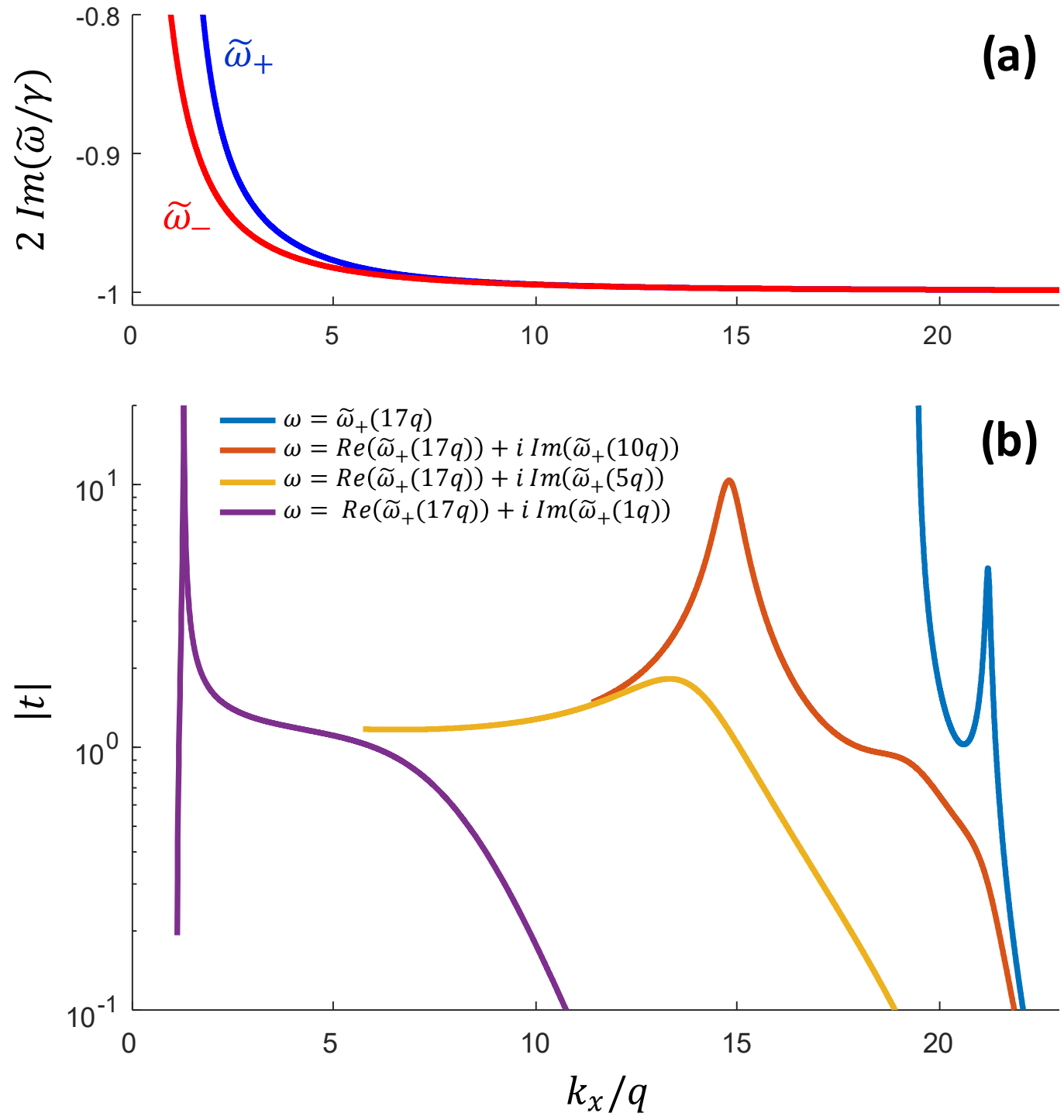

**Figure S6.** **(a)** $\mathrm{Im}(\tilde{\omega}_+)$ and $\mathrm{Im}(\tilde{\omega}_-)$ as a function of the spatial frequency $k_x$. **(b)** $|t|$ for four different complex frequency illuminations, where the real part is fixed at $\mathrm{Re}\big(\tilde{\omega}_+(17q)\big)$. The damping values of the imaginary parts are progressively decreased as: $|\mathrm{Im}(\omega)| = \big|\mathrm{Im}\big(\tilde{\omega}_+(17q)\big)\big| < \big|\mathrm{Im}\big(\tilde{\omega}_+(10q)\big)\big| < \big|\mathrm{Im}\big(\tilde{\omega}_+(5q)\big)\big| < \big|\mathrm{Im}\big(\tilde{\omega}_+(1q)\big)\big|$ . Only the spatial frequency ranges where the steady-state regime is observable are included. At low $k_x$ values, the transient responses persist longer than the steady-state responses.

# 6. Resilience to material loss

The optimal frequency, $\ddot{\omega}_2$, is very close to the arithmetic mean of $\tilde{\omega}_+(18q)$ and $\tilde{\omega}_-(18q)$, which we denote as $\ddot{\omega} = [\tilde{\omega}_+(18q) + \tilde{\omega}_-(18q)]/2$.

This frequency demonstrates resilience to material losses in the lens or background medium, as illustrated by the three curves in Fig. S7. These curves represent transmission levels under varying degrees of material loss. It is noteworthy that the maximum value, $K_x = 19q$, remains unaffected by changes in material loss. This value is solely influenced by the lens thickness ($d$) owing to the decaying exponential in Eq. (5). Increasing the image distance $(2d)$ between the input and output planes would drastically reduce $K_x$.

It is important to note that the optimal frequency $\ddot{\omega}$ varies from one curve to another since the surface polariton branches change with material parameters.

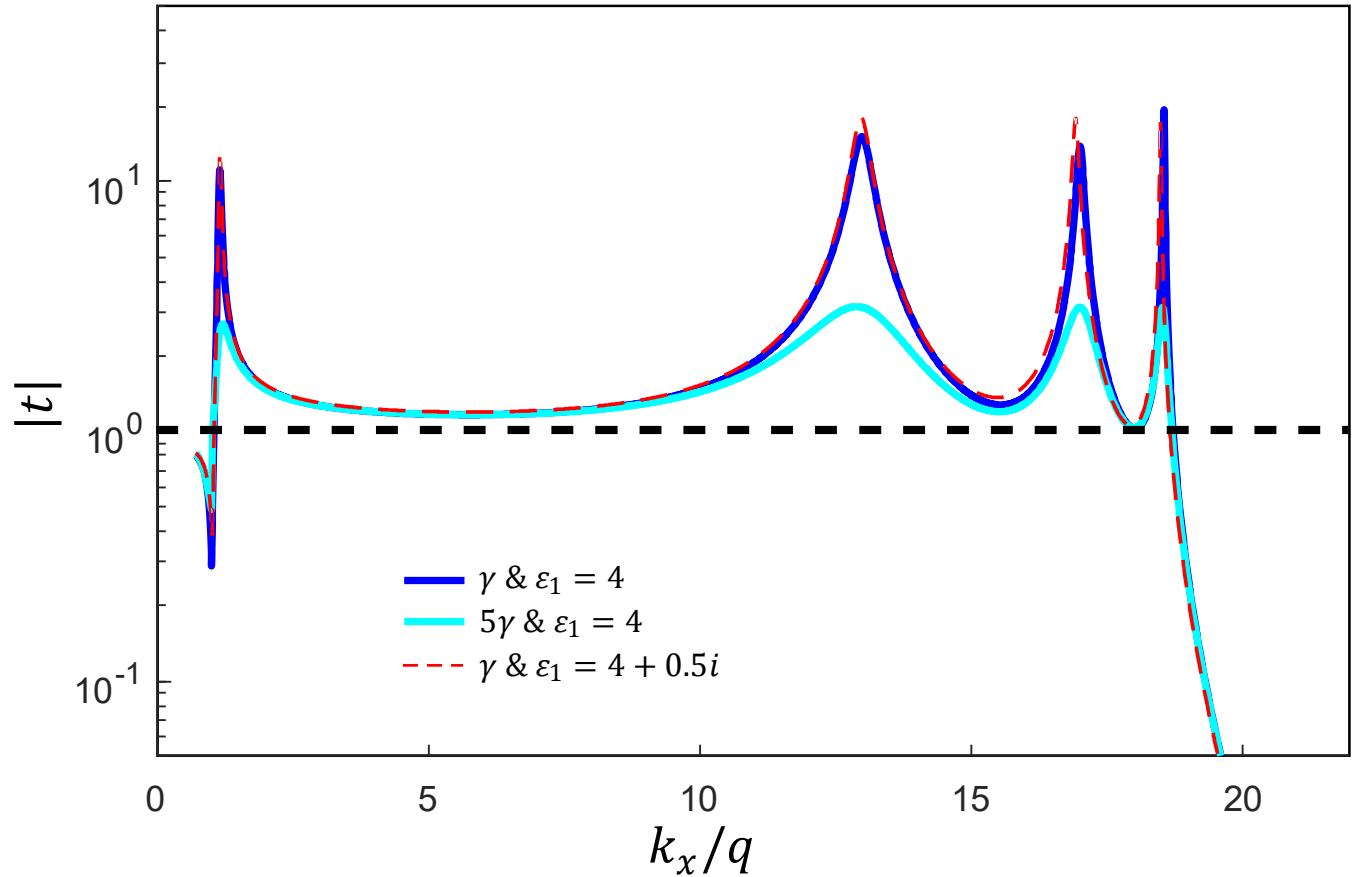

**Figure S7.** Equation (5) is generic and selecting $\omega = [\tilde{\omega}_+(18q) + \tilde{\omega}_-(18q)]/2$ as the optimal frequency is quite independent of material parameters (not on geometrical parameters like the lens thickness). The blue curve corresponds to the case presented in the main text, using a damping rate $\gamma = 0.056 \times 10^{14}$ rad/s. The cyan curve represents a scenario with increased damping, where γ is scaled up by a factor of five ($\gamma \rightarrow 5\gamma$). The red dashed curve holds for a lossy background ($\varepsilon_1 = 4 + 0.5i$), five times more absorptive than the typical infrared loss of $SiO_2$ at $\lambda = 11$ µm. All curves are computed with the $2 \times 2$ matrix formalism.

## 7. Imaging performance: $z$-component of the field

The imaging performance of the superlens is shown for the $x$-component of the electric field. Figure S7 shows the $z$-component of the electric field. Similar performance is found for both components of the electric field. The near-field computational results are obtained with the freeware [12].

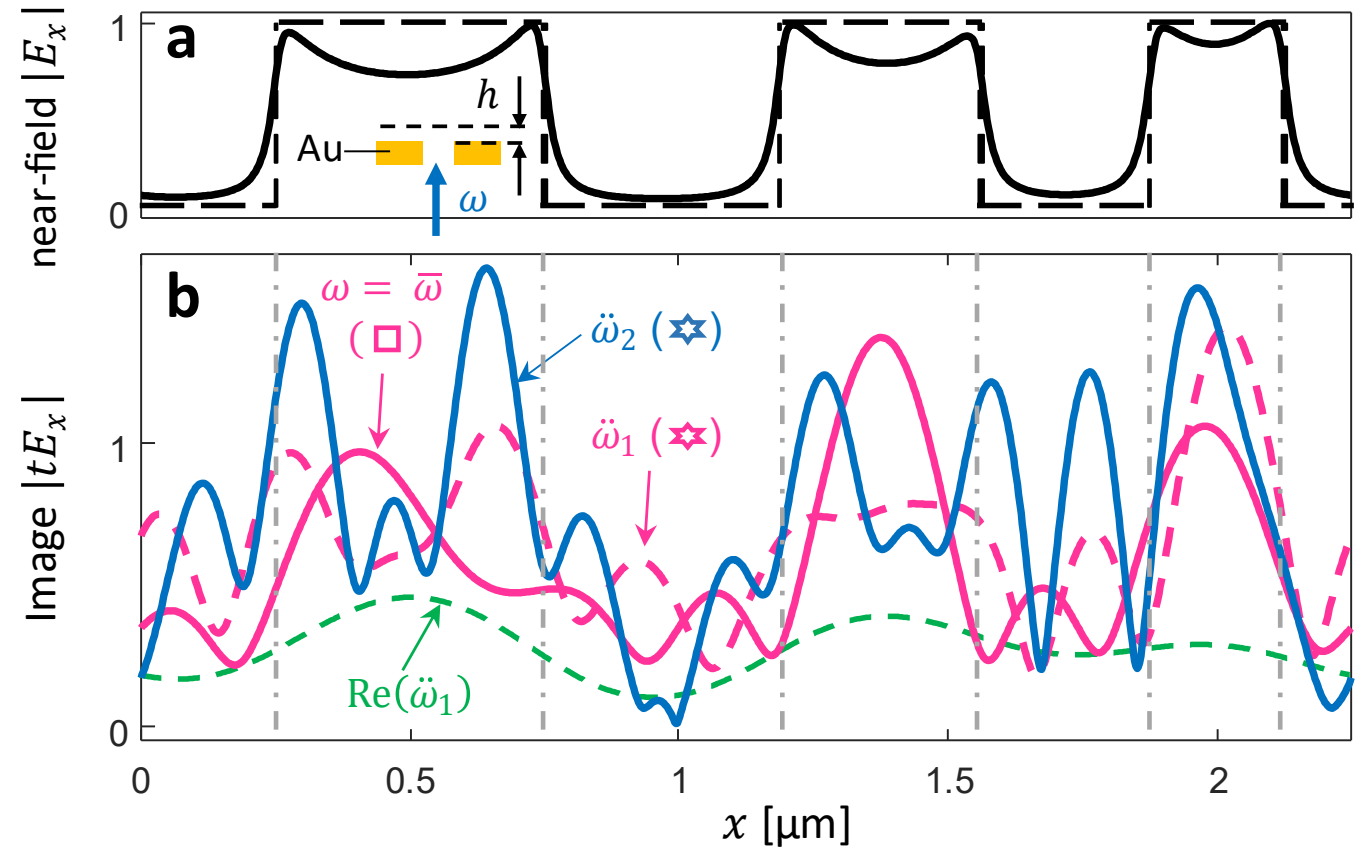

**Figure S8.** Same as in Fig. 3 for the $z$-component of the electric field. All curves are normalized by the maximum of $|E_z|$.